\documentclass[english,aps,prl,reprint,aps,superscriptaddress,floatfix]{revtex4-2}
\usepackage{bm}
\usepackage{amsmath}
\usepackage{amssymb}
\usepackage{graphicx}
\usepackage[colorlinks=true,linkcolor=blue,citecolor=blue,urlcolor=blue]{hyperref}
\usepackage{siunitx}
\usepackage{color}
\usepackage{float}
\usepackage{enumitem}
\usepackage{amsthm}
\usepackage{physics}
\usepackage{extarrows}

\begin{document}

\title{Weak localization as a probe of intervalley coherence in graphene multilayers}
\author{Nemin Wei}
\affiliation{Department of Physics, University of Texas at Austin, Austin TX 78712}
\affiliation{Department of Physics, Yale University, New Haven, CT 06520}
\author{Yongxin Zeng}
\affiliation{Department of Physics, University of Texas at Austin, Austin TX 78712}
\affiliation{Department of Physics, Columbia University, New York, NY 10027}
\author{A. H. MacDonald}
\affiliation{Department of Physics, University of Texas at Austin, Austin TX 78712}

\date{\today} 

\begin{abstract}
Spontaneous intervalley coherence is suspected in several different graphene multilayer systems,
but is difficult to confirm because of a paucity of convenient experimental signatures. 
Here we suggest that magneto-conductance features associated with 
quantum corrections to Drude conductivity can serve as a smoking gun for 
intervalley coherence that does not break time-reversal symmetry.  In this class of ordered multilayer  
quantum transport corrections can produce weak localization or weak antilocalization, 
depending on whether the valley order belongs to the orthogonal or symplectic symmetry class.  
Our analysis motivates low-temperature weak-field magnetoresistance measurements 
in graphene multilayers in which time-reversal invariant intervalley coherent order is conjectured.
\end{abstract}

\maketitle

{\em Introduction}--- Many two-dimensional (2D) hexagonal-lattice materials, including graphene multilayers
and many monolayer and bilayer group-IV transition metal dichalcogenides, have band energy extrema at one of 
two inequivalent corners of the triangular lattice  Brillouin zone. 
The valleys surrounding these k-points are related by time-reversal symmetry and their quasiparticles 
are largely responsible for the extraordinary electrical and optical properties of these materials. 
Because their large separation in momentum space suppresses intervalley scattering, it is convenient to regard valley as a 
two-state pseudospin degree of freedom. In recent years, exotic spin and valley pseudospin order
has been discovered in a variety of strongly correlated electron platforms,
including moir\'{e} materials and quantum Hall systems
\cite{lu2019superconductors,sharpe2019emergent,serlin2020intrinsic,tao2022valleycoherent,nuckolls2023quantum,kim2023imaging,li2019scanning,liu2022visualizing,coissard2022imaging,farahi2023broken}.

Probing valley order in graphene multilayers is in general nontrivial. 
Ising-like valley polarized states break time-reversal and are usually identified by anomalous Hall effects \cite{sharpe2019emergent,serlin2020intrinsic,chen2020tunable,lin2022spin,kuiri2022spontaneous,he2021competing}.
Time-reversal invariant intervalley coherent (IVC) states, hypothesized in various systems \cite{he2021competing,zhang2023enhanced,kwan2022kekule,lu2022correlated,xie2023flavor,das2023quartermetal}, do not have such obvious transport signatures. 
For instance, IVC metals are leading candidates for the partially isospin polarized phases adjacent to 
superconducting states in rohombohedral ABC trilayer graphene \cite{zhou2021superconductivity,huang2022spin,chatterjee2021intervalley}. A convenient widely applicable 
procedure for identifying intervalley coherence whenever it occurs might help unravel many of the mysteries of 
these materials, for example by shedding light on the mechanism for graphene-based superconductivity \cite{cao2018unconventional,su2023superconductivity,khalaf2021charged,chatterjee2022skyrmion,kozii2020superconductivity,chatterjee2021intervalley,christos2023nodal,dong2023signatures,dong2023superconductivity}.

Scanning tunneling spectroscopy (STM) is currently the main tool to detect IVC order in graphene.
IVC order normally yields $\sqrt{3}\times\sqrt{3}$ Kekul\'{e} patterns that triple the graphene unit cell area in 
atomic-scale STM images, as demonstrated in monolayer graphene under a strong magnetic field and in 
magic angle twisted bilayer and trilayer graphene \cite{li2019scanning,liu2022visualizing,coissard2022imaging,farahi2023broken,nuckolls2023quantum,kim2023imaging}.  
However, not all IVC order yields Kekul\'{e}-signals, notably the Kramers-IVC (K-IVC) order predicted by mean-field theories in twisted bilayer graphene \cite{calugaru2021spetroscopy, hong2021detecting,bultinck2020ground,lian2021twisted,zhang2020correlated,song2022magic}.
Moreover, 
strongly correlated states that might be IVC often appear at low charge densities and large out-of-plane displacement fields \cite{zhang2023enhanced,chatterjee2021intervalley}.  These states therefore appear only in 
dual-gated devices that are incompatible with 
STM.  A complementary probe of intervalley coherence is therefore highly desirable \cite{thomson2022gate,xie2023gate}.

In conventional weakly disordered metals with orbital time-reversal symmetry, backscattering off disorder is enhanced by constructive interference between time-reversed quasiparticle paths, yielding a negative quantum correction to the classical Drude conductivity. 
This effect is referred to as weak-localization (WL) \cite{lee1985disordered}. Strong spin-orbit coupling that mixes the 
two spin flavors makes the interference destructive and leads to weak-antilocalization (WAL) \cite{hikami1980spin,bergmann1984weak, iordanskii1994weak, knap1996weak,araki2014weak}. 
These interference effects are readily identified in magnetoconductance measurements \cite{altshuler1980magnetoresistance}, and are highly susceptible to time-reversal symmetry breaking at low temperatures. 

In this work, we propose that quantum interference corrections to conductivity might serve as a transport signature
of time-reversal invariant IVC order in metallic graphene systems. 
We first observe that when quasiparticles have strictly conserved valley numbers, the energy difference between two Bloch states at opposite momenta of the same valley, produced by trigonal warping, eliminates enhanced backscattering \cite{mccann2006weak}. 
WL corrections to the conductivity are therefore strongly suppressed in a valley conserving system. 
IVC order can mix valleys, however, and enable interference between time-reversed trajectories of mean-field quasiparticles. When spin is neglected, this interference leads to WL if the valley order preserves orbital time-reversal $\mathcal{T}_{O}$, which takes the complex conjugate of real-space wave functions and therefore flips the valley quantum number $\tau^z(=\pm 1)$ of low-energy electrons. The conjectured K-IVC states do not have 
$\mathcal{T}_{O}$ symmetry, and instead features a generalized form of time-reversal symmetry, 
$\mathcal{T}_{K}=-i\tau^z\mathcal{T}_{O}$, which is a combination of $\mathcal{T}_{O}$ and valley U(1). We will refer to this as (spinless) Kramers time-reversal symmetry since $\mathcal{T}_{K}^2=-1$, 
and show that it gives rise to WAL.  Thus low-temperature magnetoresistance measurements not only signal intervalley coherence,
but also distinguish between the two types of generalized time-reversal symmetry in conjectured IVC states.
Below we first illustrate this idea with two concrete examples by examining WL in strong-displacement-field ABC trilayer 
graphene with IVC order and WAL in Bernal stacked bilayer graphene with K-IVC order and then    
comment on the role of intervalley disorder scattering and beyond-mean-field electron-electron interactions.

\begin{figure}
    \includegraphics[width=1\columnwidth]{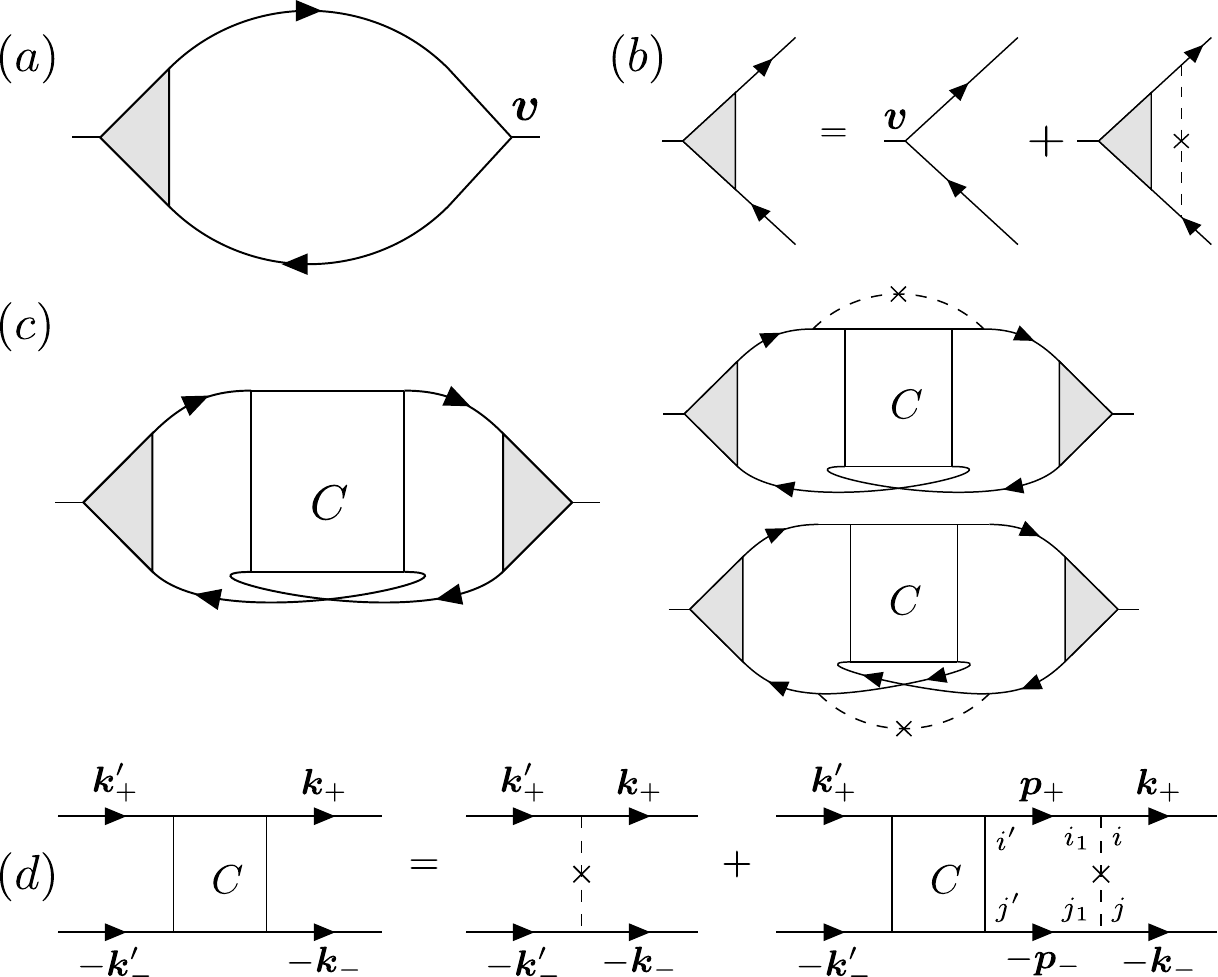}
    \caption{(a) Feynman diagram for the Drude conductivity with vertex correction (b). $\hat{\bm v}_{\bm k}=\partial \hat{H}^{0}/\partial \bm k$ is the velocity operator. (c) Hikami boxes relating the conductivity correction to the Cooperon propagator. The left diagram is the bare Hikami box and the right two diagrams are dressed Hikami boxes. Solid lines represent disorder averaged retarded and advanced Greens functions $\hat{G}^{R,A}$; 
    dashed lines represent impurity-averaged disorder potentials $\check{U}$. (d) The ladder diagram sum of the Cooperon propagator.}
    \label{fig:cooperon}
\end{figure}

{\em WL correction to conductivity}--- 
We consider mean-field Hamiltonians of the form $H_{\bm{k}i,\bm{k}'j}=H_{ij}^{0}(\bm{k})\delta_{\bm{k},\bm{k}'}+V_{\bm{k}i,\bm{k}'j}$, where $\hat{H}^{0}$ is translationally invariant and therefore preserves quasimomentum $\bm{k}$,
$\hat{V}$ is a disorder potential with zero spatial average ($\langle\hat{V}\rangle_{\text{dis}}=0$), and 
$i,j$ are conflated labels for other degrees of freedom including valley.
Figs.~\ref{fig:cooperon}a and c summarize the Feynman diagrams that contribute to the Drude conductivity $\sigma_{D}$ and its quantum corrections \cite{mccann2006weak}. Physically the quantum correction is due to interference between time-reversed multiple disorder scattering paths, as shown in Fig.~\ref{fig:cooperon}d. The Cooperon ladder diagram sum $\check{C}$ 
satisfies a Bethe-Salpeter equation,
$\check{K}\cdot\check{C}=\check{U}$, with kernel matrix
\begin{equation}
\begin{split}
        K_{\bm k ij,\bm p i'j'}(\bm{q},\omega)=& \delta_{\bm k,\bm p}\delta_{i,i'}\delta_{j,j'}-\sum_{i_1j_1}U_{\bm{k}ij,\bm{p}i_{1}j_{1}}(\bm q)\\
        &\times G_{i_1i'}^{R}\left(\bm p_{+},\epsilon+\omega \right) G^{A}_{j_1j'}\left(-\bm p_{-},\epsilon\right).
\end{split}
\end{equation}
Here, we introduced the disorder correlator represented by a dashed line in Fig.~\ref{fig:cooperon}, $U_{\bm{k}ij,\bm{p}i_{1}j_{1}}(\bm q)=\langle V_{\bm{k}_{+}i,\bm{p}_{+}i_{1}}V_{-\bm{k}_{-}j,-\bm{p}_{-}j_{1}}\rangle_{\text{dis}}$, and defined $\bm{p}_{\pm}\equiv \bm{p}\pm\bm{q}/2$ \cite{wolfle1984anisotropic,lee1985disordered}.
The disorder-averaged Green's functions $\hat{G}^{R,A}(\bm k, \epsilon)=[\epsilon \hat{1}-\hat{H}^{0}(\bm k)-\hat{\Sigma}^{R,A}(\bm k)]^{-1}$ and the self-energies $\hat{\Sigma}^{R,A}$ can be calculated within the self-consistent Born approximation. Small eigenvalues $\lambda_n$ of the kernel $\check{K}$ yield large contribution to the Cooperon matrix $\check{C}=\check{K}^{-1}\cdot\check{U}$ and hence large conductivity corrections.  Channels for which $\lambda_n\rightarrow 1$ do not make 
observable contributions to the conductivity. 

Particle number conservation and generalized time-reversal symmetry $\mathcal{T}=U_{T}\mathcal{K}$ ($U_{T}$ is a unitary matrix and $\mathcal{K}$ is complex conjugation) together ensure that $\check{K}(\bm{q}=0,\omega=0)$ has one
\textit{zero} eigenvalue \cite{wolfle1984anisotropic}.
For the multi-band circumstance of interest to us, the corresponding eigenfunction is \footnote{See Supplementary Material for (a) general properties of the Bethe-Salpeter kernel and Cooperon matrices and the derivation of the conductance corrections in multi-band systems, (b) calculation details for the conductance corrections in the IVC phase of ABC trilayer graphene and for Cooperon gaps in a model of AB-stacked Bernal bilayer graphene with K-IVC order.}, 
\begin{equation}\label{eq:cooperon}
    \phi_{\bm{k}ij}^{0}=\left[\mathcal{T}\left(\hat{\Sigma}^{R}(\bm k)-\hat{\Sigma}^{A}(\bm k)\right)\right]_{ji}.
\end{equation}

If the system does not obey other conservation laws, this will be the only gapless eigenmode and give rise to a singular Cooperon matrix at small $q,\omega$, ${C_{\bm{k}ij,\bm{k}'i'j'}(\bm{q},\omega)\approx \phi_{\bm{k}ij}^{0}(\phi_{\bm{k}'i'j'}^{0})^{*}/2\pi\gamma(\bm{q}\cdot D\cdot\bm{q}-i\omega)}$,
where $D=\sigma_{D}/e^2\gamma$ is the diffusion tensor and the single-particle density of states $\gamma=-\sum_{\bm{k}}\Im\text{tr}(\hat{G}^{R})/\pi\Omega$ must be positive \cite{finkel2010disordered}.
The WL correction to the conductivity $\delta\sigma$ is the sum of the three diagrams in Fig.~\ref{fig:cooperon}c. 
Neglecting all Cooperon modes except the gapless one leads to \footnotemark[1]
\begin{equation}\label{eq:dsigma}
    \delta\sigma^{\alpha\beta}\approx -\frac{se^2}{2\pi^2h}D^{\alpha\beta}\int\frac{d^{2} q}{\bm{q}\cdot D\cdot\bm{q}-i\omega+\tau_{\phi}^{-1}},\ 
\end{equation}
where $s=\mathcal{T}^2=\pm 1$ belong to orthogonal and symplectic symmetry classes, respectively \cite{evers2008anderson,zirnbauer2011symmetry,dyson1962statistical}, and we have introduced a decoherence rate $\tau_{\phi}^{-1}$, 
which accounts for inelastic scattering and cuts off the logarithmic divergence of $\delta\sigma$ at long wavelength. 

In the absence of broken symmetries, graphene systems with trigonally warped energy bands have a U(1) valley-number 
conservation symmetry that leads to one and only one additional gapless Cooperon. Normal metallic phases of graphene systems therefore have two gapless Cooperon eigenmodes given by Eq.~\eqref{eq:cooperon} with $\mathcal{T}=\mathcal{T}_{O}$ and $\mathcal{T}_{K}$, respectively. Since $\mathcal{T}_{O}^2=-\mathcal{T}_{K}^2=1$, Eq.~\eqref{eq:dsigma} implies that these two modes produce opposite corrections to conductivity; there is no WL or WAL in valley-conserving graphene systems.
If IVC order breaks the valley symmetry spontaneously and violates only $\mathcal{T}_{O}$ or only $\mathcal{T}_{K}$
but not both symmetries, one Cooperon will become gapped.
The remaining gapless Cooperon will induce WL when the remaining symmetry is $\mathcal{T}_{O}$
and WAL when the remaining symmetry is $\mathcal{T}_{K}$.

{\em ABC Graphene}--- We contrast the WL effects of normal and IVC metals 
using rhombohedral ABC trilayer graphene in the large displacement field \cite{zhou2021superconductivity} limit as 
an example.  Because the displacement field polarizes band-edge carriers onto a single sublattice on one of the outer layers, the low-energy Hamiltonian is similar to a two-dimensional electron gas model because graphene's sublattice 
degree of freedom is lost. The Hamiltonian can be written 
quite generally in the form
\begin{equation}\label{eq:H0}
    \hat{H}^{0}(\bm k)=\bar{\epsilon}_{\bm k}\tau^{0}+\frac{1}{2}\bm \Delta_{\bm k}\cdot\bm\tau,
\end{equation}
where $\tau^{0}$ is the identity matrix and $\bm\tau$ are three Pauli matrices in the valley space. $\bm\Delta_{\bm k}$ is a
valley pseudospin splitting field; the $x$ and $y$ components of $\bm\Delta$ break the valley $U(1)$ symmetry.
For simplicity, we take a random scalar potential $\hat{V}=u_{0}(\bm r)\tau^{0}$ with short-range correlation $\langle u_0(\bm r)u_{0}(\bm r')\rangle_{\text{dis}}=u_{0}^2\delta(\bm r-\bm r')$ as the dominant type of disorder. Provided that band mixing by disorder scattering is negligible, Green's functions take the form,
\begin{equation}\label{eq:GRA}
    \hat{G}^{R/A}(\bm k,\epsilon) = \sum_{l=\pm}\frac{|\bm{k}l\rangle\langle \bm{k}l|}{\epsilon-\epsilon_{\bm k l}\pm i\hbar/2\tau_{\bm k l}},
\end{equation}
where $\epsilon_{\bm k l}$ is the quasiparticle energy and $\tau_{\bm k l}$ is the relaxation time of the valley state $|\bm{k}l\rangle$ at momentum $\bm{k}$ in the band $l$
. Let us define $n_{\bm{k} l}^{\mu}=\langle\bm{k}l|\tau^{\mu}|\bm{k}l\rangle$, the Fermi surface average 
$\langle n^{\mu}\rangle=\sum_{\bm{k}, l}\delta(\epsilon-\epsilon_{\bm{k} l})n_{\bm{k}l}^{\mu}/\gamma\Omega$, and $\tau_{0}= (\pi u_0^2\gamma)^{-1}$. From the self-consistent equation,
\begin{equation}
\hat{\Sigma}^{R/A} = \frac{u_0^2}{\Omega}\sum_{\bm k}\hat{G}^{R/A}(\bm{k},\epsilon) = \mp \frac{i}{2\tau_0}\sum_{\mu=0,x,y,z}\langle n^{\mu}\rangle \tau^{\mu},    
\end{equation}
we obtain $\tau_{\bm{k} l}^{-1}= -2\Im\langle\bm{k}l|\hat{\Sigma}^{R}|\bm{k}l\rangle = \tau_{0}^{-1}\sum_{\mu}n_{\bm{k} l}^{\mu}\langle n^{\mu}\rangle$.
It is convenient to solve the Bethe-Salpeter equation in the valley triplet/singlet basis 
$(\tau^{\mu}\tau^{x})_{i,j}/\sqrt{2}$ where $\nu=z$ for valley singlets and $\nu=0,x,y$ for valley 
triplets.  Since  we expect solutions that 
depend only on $\bm q,\omega$ and not on $\bm{k},\bm{k}'$, it follows that 
\begin{equation}\label{eq:ansatz}
    C_{\bm k ij,\bm{k'}i'j'}(\bm q,\omega) = \frac{1}{2}\sum_{\mu\nu}(\tau^{\mu}\tau^{x})_{ij}C_{\bm q,\omega}^{\mu\nu}(\tau^{\nu}\tau^{x})_{i'j'}^{*},
\end{equation}
where $C_{\bm q,\omega}^{\delta\nu}$ solves the two-particle valley-state matrix equation 
$\sum_{\delta}K_{\bm q,\omega}^{\mu\delta}C_{\bm q,\omega}^{\delta\nu}=u_0^2\delta^{\mu\nu}$ with
\begin{equation} \label{eq:Kmunu_q}
    K_{\bm q,\omega}^{\mu\nu} = \delta^{\mu\nu}-\frac{u_0^2}{2\Omega}\sum_{\bm k}\text{tr}\!\left[\hat{G}_{\bm k_{+},\epsilon+\omega}^{R}\tau^{\mu}\tau^{x}(\hat{G}_{-\bm k_{-},\epsilon}^{A})^{\text{t}}\tau^{x}\tau^{\nu}\right].
\end{equation}

For the valley paramagnetic state, $\hat{H}^{0}$ is constructed from the non-interacting band dispersion of the tight-binding Hamiltonian \cite{cea2022superconductivity,zhang2010band}, with $\bm\Delta_{\bm k} = (\epsilon_{\bm{k},K}-\epsilon_{\bm{k},-K})\hat{z}$ generating trigonal warping in the bandstructure. In this case we can use the valley labels $\pm K$ as band indices $l$, and 
obtain the Bloch vector $\bm{n}_{\bm{k}\pm}=\pm \hat{z}$ and relaxation time $\tau_{\bm k l}=\tau_{0}$. Owing to valley-number conservation, the kernel can be decomposed into intervalley $(\mu=0,z)$ and intravalley $(\mu=x,y)$ sectors. 
In the intervalley sectors, $K_{0,0}^{\mu\nu}=0$ and the two intervalley Cooperons are gapless. However, their contributions to the conductance corrections cancel as explained in the previous section. 
The intravalley Cooperons acquire a gap, 
\begin{equation}
    \lambda_{\text{intra}}=\frac{2}{\gamma}\int \frac{d^2k}{(2\pi)^2} \delta(\epsilon-\epsilon_{\bm k K})\frac{
    (\Delta_{\bm k}^{z}\tau_0/\hbar)^2}{1+(\Delta_{\bm k}^{z}\tau_0/\hbar)^2}.
\end{equation}
$\lambda_{\text{intra}}\approx 1$ \footnotemark[1] and the WL effect is completely suppressed when the trigonal warping energy scale is 
larger than the disorder broadening scale $\hbar/\tau_{0}$. This condition is normally satisfied in high quality graphene samples.

\begin{figure}
    \centering
    \includegraphics[width=\linewidth]{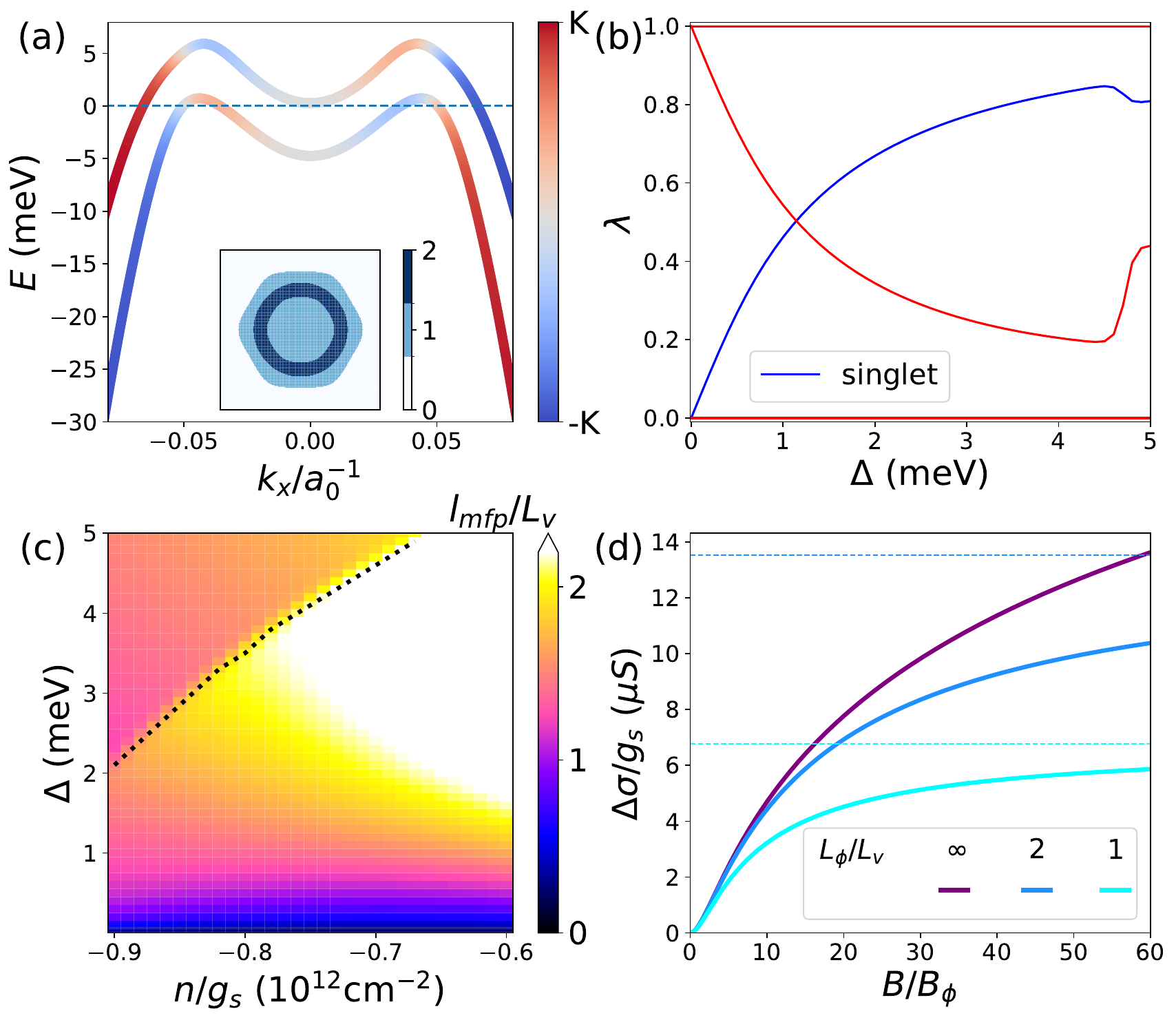}
    \caption{Cooperons and magnetoconductance in ABC trilayer graphene. (a) The electronic structure of a partially isospin polarized state with valley-XY exchange field $\Delta=5$\si{meV} and potential difference between two outer layers $U=40$\si{meV}. The dashed line is the Fermi level at hole density per spin $-n/g_s=7\times10^{11}$\si{\centi\meter^{-2}}. The inset plots the hole occupation number in reciprocal space. (b) The eigenvalues $\lambda$ of $\hat{K}_{0,0}$ \textit{vs.} the exchange field $\Delta$. The gapless Cooperon is protected by $\mathcal{T}_{O}$ symmetry, whereas the valley singlet mode (blue curve)
    is gapped by IVC order. (c) The ratio of the mean free path $l_{mfp}$ and the valley relaxation length $L_v$. Without IVC, $L_v\gg l_{mfp}$ is limited by intervalley scattering induced by atomically sharp defects or sample boundaries not included in this calculation. In contrast, $L_v\lesssim l_{mfp}$ for $\Delta$ as weak as \SI{1}{meV}. The white dashed line marks a Lifshitz transition, above which the Fermi level moves below the $\bm k=0$ local minimum of the upper band and the electron-like Fermi pocket of the upper band vanishes. (d) Out-of-plane magnetoconductance per spin $\Delta\sigma(B)/g_s$, which saturates to $e^2/2\pi h\times \ln(1+2L_{\phi}^2/L_{v}^2)$ as marked by dashed lines.} 
    \label{fig:trilayer}
\end{figure}

IVC metals are generated by finite $\Delta_{\bm k}^{x,y}$. Since the precise form of $\Delta_{\bm{k}}^{x,y}$ plays no essential role in WL properties we will
assume that $\Delta_{\bm k}^x=\Delta$ and $\Delta_{\bm k}^{y}=0$, which preserves $\mathcal{T}_{O}=\tau^{x}\mathcal{K}$ symmetry. Fig.~\ref{fig:trilayer}b depicts a quasiparticle band structure $\epsilon_{\bm k \pm}=\bar{\epsilon}_{\bm k}\pm |\bm{\Delta_{k}}|/2$ in the IVC phase in which the color encodes the $z$ component of the Bloch vector $\bm{n}_{\bm{k}l}=l\bm{\Delta_k}/|\bm{\Delta_k}|$. When the band splitting near the Fermi level 
is much larger than the disorder broadening ($\Delta\gg \hbar/\tau_{0}$), Eq.~\eqref{eq:GRA} is valid.
In this limit we can drop interband interference. The kernel in Eq.~\eqref{eq:Kmunu_q} then simplifies to
\begin{align}
    K_{\bm q,\omega}^{\mu\nu} \approx \delta^{\mu\nu}- \frac{1}{\tau_0}\left\langle \frac{n^{\mu}n^{\nu}\tau}{1+i(\bm q \cdot \bm v -\omega)\tau}\right\rangle \label{eq:Kmunu},
\end{align}
where we have used $\mathcal{T}_{O}$ symmetry $\tau^{x}(\hat{G}_{-\bm{k},\epsilon}^{A})^{\text{t}}\tau^{x}=\hat{G}_{\bm{k},\epsilon}^{A}$.
Among the four eigenvalues of $\hat{K}_{0,0}$ in Fig.~\ref{fig:trilayer}b, there is one zero eigenvalue associated with the eigenvector $\langle n^{\mu}\rangle$ and another 
$\lambda_s=1-\langle\tau (n^z)^2\rangle/\tau_0$, associated with the eigenvector $\delta^{\mu,z}$, 
which drops to zero as the IVC order vanishes and $|n_{\bm k l}^{z}|= 1$. 
The former is the gapless Cooperon mode $\phi_{\bm{k}ij}^{0}\propto\sum_{\mu}\langle n^{\mu} \rangle(\tau^{\mu}\tau^{x})_{ij}$, expected from the general formula Eq.~\eqref{eq:cooperon}, and the latter is the valley singlet mode $\phi_{\bm{k}ij}^{s}\propto i\tau_{ij}^{y}$. 
Due to $C_{3z}$ symmetry, $K_{\bm{q}, 0}^{z\mu} = 0$ ($\mu=0,x,y$) \footnote{In the Taylor series of $K_{\bm{q},\omega}^{z\mu}$, odd orders in $\bm q$ must vanish because they cannot be $C_{3z}$ invariant, and coefficients of even orders in $\bm q$ equal zero under time-reversal symmetry because $n_{\bm k}^{z}=-n_{-\bm k}^{z},n_{\bm k}^{x/y}=n_{-\bm k}^{x/y},\bm{v}_{\bm k}=-\bm{v}_{-\bm k}$.} and these two modes are decoupled.  An approximate expression for $C^{\mu\nu}$ can be constructed from these small eigenvalues and the associated eigenvectors: 
\begin{equation} \label{eq:cqmunu}
    C_{\bm q,0}^{\mu\nu} \approx \frac{1}{2\pi\gamma \tau_0^2}\left(\frac{\langle n^{\mu}\rangle\langle n^{\nu}\rangle}{Dq^2}+\frac{\delta^{\mu,z}\delta^{\nu,z}}{D_sq^2+\lambda_s\tau_0^{-1}}\right),
\end{equation}
where $D_{s}=\langle v^2\tau^{3}(n^z)^2\rangle/2\tau_{0}^2$ differs from the diffusion constant $D$ in the presence of IVC order. 
The first term is responsible for the singular WL correction Eq.~\eqref{eq:dsigma}, while the second Cooperon generates an opposite but less singular conductance correction \footnotemark[1]. Both diminish in a weak out-of-plane magnetic field, leading to the d.c. magnetoconductance $\sigma(B)-\sigma(0)\equiv\Delta \sigma(B)$\footnotemark[1],
\begin{align}\label{eq:dsigma_tri}
    &\Delta\sigma(B) \approx \frac{g_s e^2}{2\pi h }\left[F\left(\frac{B}{B_{\phi}}\right) - F\left(\frac{B}{B_{\phi}+2B_{v}}\right)\right],\\
    &F(x)= \ln{x}+\psi\left(\frac{1}{2}+\frac{1}{x}\right),\ \ B_{\phi,v} \equiv \frac{\hbar}{4eL_{\phi,v}^2},\notag
\end{align}
with the dephasing length $L_{\phi}=\sqrt{D\tau_{\phi}}$ and the valley relaxation length $L_{v}=\sqrt{2D_s\tau_0/\lambda_s}$. $L_\phi$ increases as temperature drops and can exceed $1$\si{\micro\meter}, \textit{i.e.}, implying that $B_{\phi}\lesssim 0.1$\si{mT}, 
in many graphene systems at sub-Kelvin temperatures \cite{engels2014limitations,couto2014random}. Fig.~\ref{fig:trilayer}c plots the ratio of the mean free path defined as 
$l_{mfp} \equiv \sigma_{D}^{xx}h/e^2g_sk_{F}$ with $k_{F}\equiv \sqrt{2\pi |n|/g_s}$ and $L_{v}$. 
This ratio is independent of disorder strength when $\Delta\gg \hbar/\tau_{0}$. 
Although $L_{v}$ diverges for $\Delta=0$, it decreases rapidly and becomes comparable to or shorter than $l_{mfp}$ when $\Delta$ reaches the trigonal warping energy scale $\sim 1$\si{meV}. Comparing $\Delta\sigma(B)$ at different $L_{\phi}/L_v$ in Fig.~\ref{fig:trilayer}d, we conclude that IVC order shortens $L_v$ and by doing so enhances the positive magnetoconductance.


{\em Kramers intervalley coherent order}--- It has been proposed that 
valley symmetry breaking could be present
near charge neutrality in twisted graphene multilayers with low strain, and that  
K-IVC order with $\mathcal{T}_{K}$ symmetry \cite{bultinck2020ground} is a specific possibility. 
Metallic states with K-IVC order will exhibit WAL. 
Consider for example a fictitious model of AB-stacked Bernal bilayer graphene with K-IVC order: 
\begin{align}
    \hat{H}^{0} &= \frac{k_x^2-k_y^2}{2m}\sigma^x+\frac{k_xk_y}{m}\tau^z\sigma^y +\Delta\tau^{y}\sigma^{y} + \hat{H}_{w},\\
    \hat{H}_{w} &= v_3(k_y\tau^z\sigma^{x}+k_x\sigma^{y}),
\end{align}
where the components of $\bm\sigma$ are Pauli matrices in layer space. 
The operator $\tau^y\sigma^y$, which establishes K-IVC order \cite{bultinck2020ground}, 
is an intervalley particle-hole order parameter since it couples valence band states in 
one valley with conduction band states in the other valley. $\hat{H}_w$ is responsible for trigonal warping. 
The model respects $\mathcal{T}_{K}=\tau^{y}\mathcal{K}$, but for finite $\Delta$ breaks $\mathcal{T}_{O}$ .

\begin{table}
    \centering
    \renewcommand{\arraystretch}{1.6}
    \caption{The smallest $\bm q=0$ Cooperon relaxation gaps in the four two-particle flavor-state channels. $\epsilon=\sqrt{(k_{F}^2/2m)^2+\Delta^2}$ is the Fermi energy. $\tau$ is the scattering time. $\tau_w^{-1}=2(v_3k_F)^2\tau$ is generated by trigonal warping. 
    The valley label $\xi^z=\pm 1$ is represented by $|\uparrow/\downarrow\rangle$.}
    \begin{tabular}{|c|c|c|c|c|}
        \hline
         $\lambda/\tau$ & $|\!\uparrow\downarrow\rangle-|\!\downarrow\uparrow\rangle$ & $|\!\uparrow\downarrow\rangle + |\!\downarrow\uparrow\rangle$& $|\!\uparrow\uparrow\rangle - |\!\downarrow\downarrow\rangle$ & $|\!\uparrow\uparrow\rangle + |\!\downarrow\downarrow\rangle$\\ \hline
        $\Delta\ll \epsilon$&$0$ & $\tau_w^{-1}$ & $\frac{2\Delta^2}{\epsilon^2}\tau^{-1}\!$ & $\frac{2\Delta^2}{\epsilon^2}\tau^{-1}+\tau_w^{-1}$ \\ \hline
        $\Delta\sim \epsilon$& $0$ & $2\tau_w^{-1}$ & $\tau_w^{-1}$& $\tau_w^{-1}$\\ \hline
    \end{tabular}
    \label{tab:gap_kivc}
\end{table}

To simplify the calculation of conductivity corrections
we assume that the random potential is pseudospin-independent, $\langle\hat{V}(\bm{r})\hat{V}(\bm{r}')\rangle_{\textrm{dis}} = u_0^2\tau^{0}\sigma^{0}\delta(\bm r-\bm{r}')$, and that 
trigonal warping at the Fermi level is much smaller than the scattering rate. 
The scattering rate of the conduction band electrons $\tau^{-1}=\pi u_0^2\gamma(1+\Delta^2/\epsilon^2)$ at $\epsilon=\sqrt{(k_F^2/2m)^2+\Delta^2}$. Using the approximate flavor symmetry $\xi^z=\tau^x\sigma^x$ of the Hamiltonian (weakly broken by $\hat{H}_w$),
we solved the Bethe-Salpeter equation at $\bm q=\omega=0$ \footnotemark[1].
Table~\ref{tab:gap_kivc} summarizes the resulting analytical expressions for the gaps of $\bm q=0$ Cooperons 
in different channels in two opposite limits. In both cases, the flavor singlet mode is gapless and yields WAL due to $\mathcal{T}_{K}$ symmetry. For the limit $\Delta\ll k_F^2/2m\sim \epsilon$, the third mode in the table acquires a relaxation gap $2\Delta^2/\epsilon^2\tau$. 
This mode reduces to
the gapless Cooperon associated with $\mathcal{T}_{O}$ symmetry at $\Delta=0$. In the other limit $\Delta\sim\epsilon\gg k_{F}^2/2m$, two conduction bands are polarized into $\tau^y\sigma^y=1$ and form a pseudospin degree of freedom, $\bm{\xi} \equiv (-\tau^z\sigma^{x},\tau^y,\tau^x\sigma^x)$ . The low-energy effective Hamiltonian $\hat{H}^{\text{eff}}= \bar{\epsilon}_{\bm k}\xi^{0} + v_3k_x\xi^y-v_3k_y\xi^x+\hat{V}$ describes the motion of a spin-1/2 particle with Rashba spin-orbit coupling subject to a spin-independent random potential. 
The system is known to host one gapless singlet Cooperon and three gapped triplet Cooperons and exhibit WAL \cite{iordanskii1994weak,knap1996weak,araki2014weak}.


{\em Discussion}---
In this work, we point out that under some circumstances (see below)
intervalley coherence (IVC) order in graphene multilayers will induce 
quantum interference corrections to conductivity, motivating  
low-temperature weak-field magnetoresistance measurements when this order is suspected.
We illustrated this idea by studying WL in ABC trilayer graphene with IVC order that is 
consistent with ordinary orbital time-reversal symmetry and WAL in a toy model of bilayer graphene with K-IVC order \cite{khalaf2021charged}, but the same principle applies to other graphene systems in which IVC order has been conjectured including ABC trilayer graphene with aligned hexagonal boron nitride \cite{chen2020tunable,patri2022trilayer}, and twisted double bilayer graphene \cite{kuiri2022spontaneous,zhang2021visualizing}. We now highlight a few limitations of our proposal.

Firstly, only time-reversal invariant IVC phases can be revealed by magnetoresistance measurements. 
If valley polarization coexists with IVC order \cite{arp2023intervalley}, it will break time-reversal symmetry, suppress WL, 
and hinder the detection of the IVC order. To see this, we notice from Eq.~\eqref{eq:Kmunu} that 
when $\epsilon_{\bm k l}-\epsilon_{-\bm k l}\gtrsim \langle\tau^{-1}\rangle$, $K^{\mu\nu}\rightarrow \delta^{\mu\nu}$, and the multiple scattering interference is destroyed.
Importantly, we have so far neglected other physical effects that break valley-number conservation, such as atomically sharp defects which conserve spin.  
These can also lead to WL and positive magnetoconductance \cite{mccann2006weak,fal2007weak,tikhonenko2008weak,electronic2011dassarma}, 
and may not always be distinguishable from spontaneous IVC.
However, given that intervalley scattering mainly occurs at the edges in the high quality graphene samples, 
we expect that the valley relaxation length $L_v$ should reach the sample dimension $L$ in the absence of 
IVC order, whereas it can (as shown in Fig.~\ref{fig:trilayer}d) be comparable to the mean free path $l_{mfp} \ll L$ in IVC phases. 
The choice between scattering and IVC interpretations can therefore decided by fitting WL data  
to Eq.~\eqref{eq:dsigma_tri}.  
In contrast, WAL cannot be generated by spin-conserving intervalley disorder scattering 
and negative magnetoconductance is therefore always a signature of K-IVC order. 
Weak intervalley scattering instead produces a small gap for the gapless Cooperon in Table~\ref{tab:gap_kivc} 
and simply reduces the WAL effect.


Last but not least, we have neglected interactions between quasiparticles 
which can also \cite{altshuler1980interaction,fukuyama1980effects,lee1985disordered} lead to singular conductance corrections. 
Because they are less sensitive to out-of-plane magnetic fields,
weak interaction effects can usually be distinguished from WL and WAL. 
(See Refs.\;\onlinecite{larkin1980reluctance,altshuler1981anomalous} for discussions of higher order effects.)
Because Cooperons do not produce singular corrections to the Hall resistance \cite{altshuler1980magnetoresistance}, interaction corrections 
can also be disentangled from WL by examining weak-field Hall data.
There are, however, circumstances in which interaction effects might be difficult to disentangle.
For example, it has been conjectured \cite{khalaf2021charged} that the charged quasiparticles in some 
graphene multilayer systems might be magnetic textures (skyrmions) that couple strongly to magnons 
and are therefore likely to have very short inelastic scattering times that reduce WL by generating a relaxation gap $\tau_{\phi}^{-1}$ for their Cooperons.
More generally, the very concept of the quasiparticles might not be well-defined as in the strange metal phase \cite{cao2020strange,jaoui2022quantum,lyu2021strange}. The applicability of our theory to these non-Fermi-liquid IVC phases warrants further study.

In spite of the above limitations, our theory could still be applicable to a variety of graphene systems. For instance, recent STM measurements support the presence of time-reversal invariant IVC order in magic-angle twisted graphene multilayers \cite{nuckolls2023quantum}. It might be promising to search for evidence of anomalous weak-field magnetoresistance, especially near band filling factors $\nu=\pm2$ in 
twisted trilayer graphene where the IVC order opens a band gap of the flat bands yet the correlated state remains conductive due to the presence of additional dispersive Dirac-like bands.
\begin{figure}[t]
    \centering
    \includegraphics[width=1\linewidth]{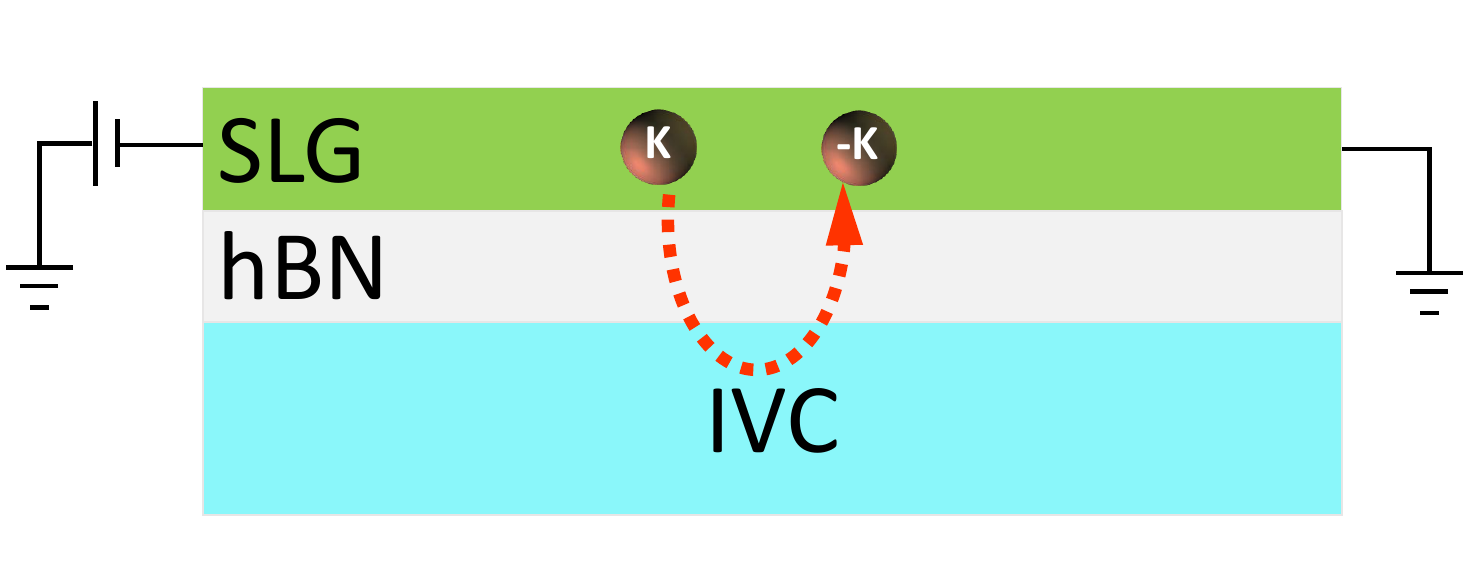}
    \caption{A heterostructure to measure IVC order in strongly correlated graphene systems. The probed system is on the bottom and is separated from a single layer graphene (SLG) by a thin layer of hexagonal boron nitride (hBN). The top SLG acquires a proximity-induced IVC order and exhibit an enhanced interference conductance correction when the bottom graphene system is tuned into the IVC regime.}
    \label{fig:device}
\end{figure}
A merit of our proposal is that it is insensitive to details of electronic structure.
IVC is, for example, expected not only in normal diffusive metal states but also in insulating and 
superconducting states and in ballistic conductors.  Fig.~\ref{fig:device} schematically illustrates  
a spectator diffusive graphene layer designed to probe for IVC order in a weakly-coupled aligned insulating state.
WL is also robust against the inevitable inhomogeneity present in real samples and against
textures in the IVC order parameters, as long as time-reversal symmetry is intact.
We  leave for future study the question of 
how intertwined spin and valley order \cite{lake2021reentrant,morissette2023dirac}, or strong spin-orbit coupling, generated by proximal transition metal dichalcogenides \cite{arora2020superconductivity}
for example, might influence quantum interference corrections to conductivity.

\begin{acknowledgements}
We gratefully acknowledge helpful interactions with A. M. Finkelstein and J.I.A. Li. We especially thank J. A. Folk for discussions that stimulated this work and for comments on an earlier version of this manuscript.
This work was supported by the U.S. Department of Energy, Office of Science, Basic Energy Sciences, under Award
\# DE-SC0022106.

\end{acknowledgements}

\bibliography{references}

\clearpage
\onecolumngrid
\setcounter{equation}{0}
\setcounter{figure}{0}
\setcounter{table}{0}
\setcounter{page}{1}
\setcounter{secnumdepth}{3}

\newcounter{proplabel}
\newtheorem{remark}{Remark}
\renewcommand{\theequation}{S\arabic{equation}}
\renewcommand{\thefigure}{S\arabic{figure}}
\renewcommand{\bibnumfmt}[1]{[S#1]}
\renewcommand{\thesection}{\Roman{section}}
\renewcommand{\thesubsection}{\arabic{subsection}}

\makeatletter
\DeclareRobustCommand{\cev}[1]{%
  \mathpalette\do@cev{#1}%
}
\newcommand{\do@cev}[2]{%
  \fix@cev{#1}{+}%
  \reflectbox{$\m@th#1\vec{\reflectbox{$\fix@cev{#1}{-}\m@th#1#2\fix@cev{#1}{+}$}}$}%
  \fix@cev{#1}{-}%
}
\newcommand{\fix@cev}[2]{%
  \ifx#1\displaystyle
    \mkern#23mu
  \else
    \ifx#1\textstyle
      \mkern#23mu
    \else
      \ifx#1\scriptstyle
        \mkern#22mu
      \else
        \mkern#22mu
      \fi
    \fi
  \fi
}
\makeatother

\begin{center}
\textbf{Supplemental materials for ``Weak localization as a probe of intervalley coherence in graphene multilayers"}
\end{center}

\section{weak localization and anti-localization effects in multi-band systems}
Let us consider a disordered multi-band system described by the following single-particle Hamiltonian,
\begin{equation}
    H=\sum_{\bm k\bm k', ij}c_{\bm k'i}^{\dagger}\left(H_{i,j}^{0}(\bm k)\delta_{\bm{k}',\bm{k}}+V_{i,j}(\bm k',\bm k)\right)c_{\bm k j},
\end{equation}
where $\bm k$ is the crystal momentum and $i,j$ label other degrees of freedom such as orbitals. $\hat{H}^{0}$ can include not only the non-interacting Hamiltonian of electrons but also the mean-field potential generated by electron-electron interactions. The random potentials $\hat{V}$ obey the following relations after the disorder average, 
\begin{equation}
\langle V_{i,j}(\bm k', \bm k)\rangle_{\text{dis}}=0 ,\quad \langle V_{i,j}(\bm{k}+\bm{q}, \bm{k})V_{i',j'}(\bm{k}',\bm{k}'+\bm{q}')\rangle_{\text{dis}} = \bra{\bm{k}'i',\bm{k}+\bm{q}i}U\ket{\bm{k}'+\bm{q} j',\bm{k}j}\delta_{\bm q, \bm q'}.
\end{equation}
On averaging the disorder, the translational symmetry is restored and the retarded/advanced Green's functions of a single particle at energy $\epsilon$, $\hat{G}^{R/A}=\langle[(\epsilon\pm i\delta)\hat{1}-\hat{H}^{0}-\hat{V}]^{-1}\rangle_{\text{dis}}$, become diagonal in the crystal momentum,
\begin{equation}
    \hat{G}^{R/A}(\bm k,\epsilon)=\left[\left(\epsilon\pm i\delta\right)\hat{1}-\hat{H}^{0}(\bm k)-\hat{\Sigma}^{R/A}(\bm k,\epsilon)\right]^{-1},\\
\end{equation}
For weak disorder potentials, we expect that $\hat{\Sigma}^{R/A}$ is small and changes slowly as the particle energy $\epsilon$ changes, $||\partial_{\epsilon}\hat{\Sigma}^{R/A}||\ll 1$. Thus for $\epsilon\approx \epsilon_0$, 
\begin{equation}\label{eq:det}
    \det\hat{G}^{R/A}(\bm k,\epsilon)^{-1} \approx \det\left(\epsilon\hat{1}-\hat{H}^{0}(\bm k)-\hat{\Sigma}^{R/A}(\bm k,\epsilon_0)\right) = \prod_{n}\left(\epsilon-\xi_{n\bm k}(\epsilon_0)\pm i\Gamma_{n\bm k}(\epsilon_0)\right),
\end{equation}
On the right hand side is the characteristic polynomial of the operator $\hat{H}^{0}+\hat{\Sigma}^{R/A}$ with $\xi_{n\bm{k}}\in \mathbb{R}$ and $\Gamma_{n\bm{k}}>0$. Because $\hat{H}^{0}$ and $\hat{\Sigma}^{R}$ should not have singularities, the poles of $\hat{G}^{R}$ come from the zeros of $\det (G^{R})^{-1}$. We can therefore interpret $\xi_{n\bm{k}}$ as the renormalized Bloch band dispersion and $\Gamma_{n\bm{k}}$ as the disorder broadening of the band $n$. Our analysis will be restricted to the weak disorder scattering limit, $\Gamma_{n\bm{k}} \ll \epsilon_{F}, \forall (n,\bm{k})$, where $\epsilon_{F}$ is the minimum distance between energy $\epsilon$ and all band energy extrema in momentum space (\textit{e.g.}, $\epsilon_{F}$ is the Fermi energy for parabolic bands). In this limit, $\hat{\Sigma}^{R/A}$ can be computed in the self-consistent Born approximation,
\begin{align}
    \Sigma_{ij}^{R/A}(\bm k,\epsilon)&=\sum_{\bm{k'}i'j'}W_{\bm{k}ij,\bm {k}'i'j'}G_{i'j'}^{R/A}(\bm k',\epsilon), \label{eq:self} \\
    &=\sum_{\bm{k'}i'j'}G_{i'j'}^{R/A}(\bm k',\epsilon)W_{\bm{k}'j'i',\bm {k}ji}, \label{eq:self2}
\end{align}
with $W_{\bm{k}ij,\bm{k'}i'j'}\equiv\bra{\bm{k}'i',\bm{k}j}U\ket{\bm{k} i,\bm{k}'j'}$. 
\begin{figure}[h]
    \includegraphics[width=1\columnwidth]{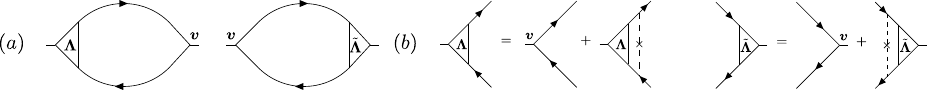}
    \caption{(a) Two equivalent Feynman diagrams for the Drude conductivity with (b) the vertex corrections.  Solid lines represent disorder averaged $G^{R,A}$ and dashed lines represent disorder. The left current vertex $\hat{\bm\Lambda}(\bm k)$ transforms to the right vertex $\tilde{\bm\Lambda}(-\bm k)$ under the time-reversal ($\textit{i.e.}$, reversing the arrow directions in the solid lines).}
    \label{fig:drude}
\end{figure}

Electric conductivity of weakly disordered systems can be calculated using the diagrammatic technique \cite{lee1985disordered}. The two Feynman diagrams in Fig.~\ref{fig:drude}a equivalently represent the Drude conductivity,
\begin{align}
    \sigma_{D}^{\alpha\beta} &= \frac{e^2}{\hbar\Omega}\int \frac{d\epsilon}{2\pi}\frac{\partial n_{F}}{\partial \epsilon}\sum_{\bm k}\tr \left(\hat{v}^{\alpha}(\bm k)\hat{G}^{R}(\bm k,\epsilon)\hat{\Lambda}^{\beta}(\bm k,\epsilon)\hat{G}^{A}(\bm k,\epsilon)\right) \label{eq:sigmad}\\
    & = \frac{e^2}{\hbar\Omega}\int \frac{d\epsilon}{2\pi}\frac{\partial n_{F}}{\partial \epsilon}\sum_{\bm k}\tr \left(\tilde{\Lambda}^{\alpha}(\bm k,\epsilon)\hat{G}^{R}(\bm k,\epsilon)\hat{v}^{\beta}(\bm k)\hat{G}^{A}(\bm k,\epsilon)\right) \label{eq:sigmad_2},
\end{align}
where $\hat{\bm{v}}=\partial \hat{H}^{0}/\partial\bm k$ is the current operator and $\hat{\bm{\Lambda}}\ (\tilde{\bm{\Lambda}})$ are the left (right) current vertex renormalized by disorder scattering (see Fig.~\ref{fig:drude}b),
\begin{equation}\label{eq:vertex_v}
    \bm \Lambda_{ij}(\bm k,\epsilon) = \bm{v}_{ij}(\bm k) + \sum_{\bm k'i'j'}W_{\bm k ij,\bm k'i'j'}\left(\hat{G}^{R}(\bm k',\epsilon)\hat{\bm \Lambda}(\bm k',\epsilon)\hat{G}^{A}(\bm k',\epsilon)\right)_{i'j'}.
\end{equation}
\begin{equation}\label{eq:vertex_vtilde}
    \tilde{\bm\Lambda}_{ij}(\bm k,\epsilon) = \bm{v}_{ij}(\bm k) + \sum_{\bm k'i'j'}\left(\hat{G}^{A}(\bm k',\epsilon)\tilde{\bm\Lambda}(\bm k',\epsilon)\hat{G}^{R}(\bm k',\epsilon)\right)_{j'i'}W_{\bm{k}' i'j',\bm{k}ij}.
\end{equation}
It can be shown that $\hat{\bm{\Lambda}}=\hat{\bm{\Lambda}}^{\dagger}$ and $\tilde{\bm{\Lambda}}=\tilde{\bm{\Lambda}}^{\dagger}$.
We emphasize that when $\hat{H}^{0}(\bm k)$ incorporates $\bm{k}$-dependent mean-field potentials generated by electron-electron interactions, $\hat{\bm v}$ differs from the bare current operator for the non-interacting system by an interaction-induced current vertex correction \cite{langer1961resistance}. 

In the presence of a generalized time-reversal symmetry $\mathcal{T}=U_{T}\mathcal{K}$, where $\mathcal{K}$ is complex conjugate and $U_T$ is a unitary matrix, the Hamiltonian and Green's functions are constrained by
\begin{align}
    &U_{T}\hat{H}^{0}(\bm k)^\text{t}U_{T}^{\dagger}=\hat{H}^{0}(-\bm k),\quad U_{T}\hat{V}({\bm{k}'},\bm{k})^\text{t}U_{T}^{\dagger}=\hat{V}({-\bm{k}},-\bm{k}'),\label{eq:trs_hv}\\
    &U_{T}\hat{G}^{R/A}(\bm k,\epsilon)^\text{t}U_{T}^{\dagger}=\hat{G}^{R/A}(-\bm k,\epsilon). \label{eq:trs_gf}
\end{align}
From Eqs.~\eqref{eq:vertex_v} and~\eqref{eq:vertex_vtilde}, the two current vertices are related by time-reversal, $U_{T}\hat{\Lambda}^{\alpha}(\bm k)^{\text{t}}U_{T}^{\dagger}=-\tilde{\Lambda}^{\alpha}(-\bm k)$. This relation, together with Eqs.~\eqref{eq:sigmad} and~\eqref{eq:sigmad_2}, implies that the Drude conductivity in a time-reversal invariant system is a symmetric tensor, $\sigma_D^{\alpha\beta}=\sigma_D^{\beta\alpha}$.

Quantum interference between time-reversed paths in multiple disorder scattering can induce weak localization (WL) and weak anti-localization (WAL) corrections to conductivity at low temperature \cite{lee1985disordered}. These interference corrections are associated with the Feynman diagram in Fig.~\ref{fig:diffuson}a, known as the Cooperon. They can be suppressed by out-of-plane magnetic fields, yielding anomalous weak-field magnetoconductance. In this section, we generalize the formalism in Ref.~\cite{wolfle1984anisotropic} to study the WL and WAL effects in multi-band systems.
\subsection{Diffusons}
\begin{figure}
    \includegraphics[width=0.6\columnwidth]{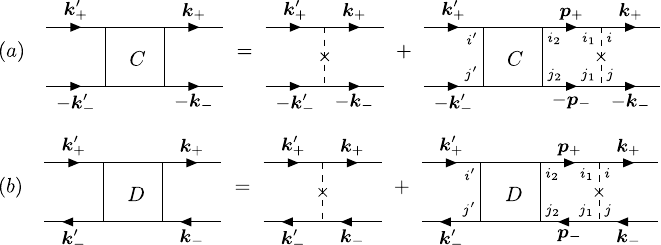}
    \caption{The ladder sum of (a) the Cooperon propagator and (b) the diffuson propagator. In a time-reversal invariant system, the Cooperon can be derived from the diffuson by a time-reversal transformation on the hole propagator in the diffuson.}
    \label{fig:diffuson}
\end{figure}
To study the Cooperon in time-reversal invariant systems, it is convenient to perform a time-reversal transformation on the bottom fermionic line in the Feynman diagram Fig.~\ref{fig:diffuson}a and obtain a particle-hole ladder diagram Fig.~\ref{fig:diffuson}b. This particle-hole propagator accounts for the diffusive density response in weakly disordered systems and is dubbed as the diffuson \cite{lee1985disordered}. The diffuson ladder sum obeys the following integral equation,
\begin{equation}\label{eq:bs_d}
     \mathcal{D}_{\bm{k}ij,\bm{k}'i'j'}(\bm{q},\omega)=W_{\bm{k}ij,\bm{k}'i'j'}(\bm q)+\sum_{\bm{p}, i_{2}, j_{2}}\sum_{i_1, j_1}W_{\bm{k}ij,\bm{p}i_{1}j_{1}}(\bm q)G_{i_{1}i_{2}}^{R}\left(\bm p_{+},\epsilon+\omega \right)G^{A}_{j_{2}j_{1}}\left(\bm p_{-},\epsilon\right)\mathcal{D}_{\bm{p}i_{2}j_{2},\bm{k}'i'j'}(\bm q,\omega),
\end{equation}
where $\bm k_{\pm}=\bm k\pm\bm q /2$. The above equation can be rewritten as $\check{K}^{D}(\bm q,\omega)\check{\mathcal{D}}=\check{W}$. The kernel $\check{K}^{D}$ reads that
\begin{align}\label{eq:KD}
    K_{\bm{k}ij,\bm{k}'i'j'}^{D}\equiv \delta_{\bm k,\bm p}\delta_{i,i'}\delta_{j,j'}-\sum_{i_1j_1}W_{\bm{k}ij,\bm{k}'i_{1}j_{1}}G_{i_1,i'}^{R}\left(\bm{k}’_{+},\epsilon+\omega\right)G_{j',j_{1}}^{A}\left(\bm{k}'_{-},\epsilon\right).
\end{align}
Here and hereafter, we neglect $\bm q-$dependence of the disorder correlator $\check{W}$, which is a good approximation at small $q$ and is exact if the disorder potential is a function of the momentum transfer, $V_{ij}(\bm{k}',\bm k)=V_{ij}(\bm k'-\bm k)$. Let us define the right and left eigenfunctions $\vec{\psi}_{\bm{k}ij}^{l}$ and $\cev{\psi}_{\bm{k}ij}^{l}$ of $\check{K}^{D}(0,0)$, $\check{K}^{D}(0,0)\vec{\psi}^{l}=\lambda_l \vec{\psi}^{l}$ and $\cev{\psi}^{l}\check{K}^{D}(0,0)=\lambda_l \cev{\psi}^{l}$, where $\lambda_{l}$'s are complex eigenvalues. $\check{K}^{D}(0,0)$ has the following properties:
\begin{enumerate}[label=\textit{Property \arabic*}. ,leftmargin=*]
    \item \label{prop1} $K_{\bm{k}ij,\bm{k}'i'j'}^{D}(0,0)=K_{\bm{k}ji,\bm{k}'j'i'}^{D}(0,0)^{*}$.
    \item \label{prop2} Eigenvalues $\lambda_l$'s are either real or in complex conjugate pairs.
    \item \label{prop4} $\check{K}^{D}(0,0)$ has a zero eigenvalue, $\lambda_{0}=0$, whose left eigenfunction is $\vec{\psi}_{\bm k}^{l}=\hat{\Sigma}^{R}(\bm k)-\hat{\Sigma}^{A}(\bm k)\equiv \Delta\hat{\Sigma}(\bm k)$, and left eigenfunction is $\cev{\psi}_{\bm k}^{l}=\Delta\hat{G}(\bm k,\epsilon)^\text{t}/\mathcal{N}$ with $\Delta\hat{G}\equiv \hat{G}^{R}-\hat{G}^{A} $ and the normalization constant $\mathcal{N}=\sum_{\bm k}\tr(\Delta G(\bm k,\epsilon)\Delta\Sigma(\bm k,\epsilon))$ such that $\cev{\psi}^{0}\cdot\vec{\psi}^{0}=1$.
\end{enumerate}

\begin{proof}
Applying the relation $W_{\bm{k}ij,\bm{k}'i'j'}=W_{\bm{k}ji,\bm{k}'j'i'}^{*}$ 
and $\hat{G}^{R}(\bm{k},\epsilon)=\hat{G}^{A}(\bm{k},\epsilon)^{\dagger}$ to Eq.~\eqref{eq:KD} yields \ref{prop1} Using this property in eigenvalue equations, we find that $\lambda_{l}^{*}$ is also an eigenvalue, $\sum_{\bm{k}'i'j'}K_{\bm{k}ij,\bm{k}'i'j'}^{D}(0,0)(\vec{\psi}_{\bm{k}'j'i'}^{l})^{*}=\lambda_{l}^{*}(\vec{\psi}_{\bm{k}ji}^{l})^{*}$, and hence establish \ref{prop2} The eigenfunctions $\vec{\psi}_{\bm{k}ij}^{l}$ and $(\vec{\psi}_{\bm{k}ji}^{l})^{*}$ are linearly independent unless $\lambda_l=\lambda_l^{*}$. 
Finally, one can show that 
\begin{equation}\label{eq:Kphi0}
    \sum_{\bm{p}i'j'}K_{\bm{k}ij,\bm{p}i'j'}^{D}(0,0)\Delta\Sigma_{i'j'}(\bm{p}) = \Delta\Sigma_{ij}(\bm{k}) - \sum_{\bm{p}i'j'}W_{\bm{k} ij,\bm{p}i'j'}\left[\hat{G}^{R}\left(\bm{p},\epsilon\right)\Delta\hat{\Sigma}\left(\bm{p},\epsilon\right) \hat{G}^{A}\left(\bm{p},\epsilon\right)\right]_{i'j'}\xlongequal[]{\text{Eq}.\eqref{eq:self}}0,
\end{equation}
and, similarly, Eq.~\eqref{eq:self2} implies that $\sum_{\bm{p}i'j'}\Delta G_{i'j'}(\bm{p},\epsilon)K_{\bm{p}i'j',\bm{k}ij}^{D}(0,0)=0$, as claimed in \ref{prop4}
\end{proof}


Note that $\vec{\psi}^{l}$'s form a complete basis and $\cev{\psi}^{l}$'s form the dual basis with the normalization condition $\sum_{\bm{k} ij}\cev{\psi}_{\bm{k} ij}^{l}\vec{\psi}_{\bm{k} ij}^{l'}=\delta_{l,l'}$. We use the eigendecomposition $K_{\bm{k}ij,\bm{k}'i'j'}^{D}(0,0)=\sum_{l}\lambda_{l} \cev{\psi}_{\bm{k} ij}^{l}\vec{\psi}_{\bm{k}' i'j'}^{l}$ to solve the integral equation Eq.~\eqref{eq:bs_d} at small $\omega$ and $q$,
\begin{equation} \label{eq:diffuson}
    \mathcal{D}_{\bm k i j, \bm{k}'i'j'}(\bm q ,\omega)\approx\sum_{\substack{l\\ \lambda_l\neq 0}}\frac{1}{\lambda_l}\vec{\psi}_{\bm{k} i j}^{l}\left(\cev{\psi}^{l}\check{W}\right)_{\bm{k}' i' j'} + \mathcal{D}_{\bm k i j, \bm{k}'i'j'}^{\text{sing}}(\bm q ,\omega),
\end{equation}
The zero modes of $\check{K}^{D}$ including $l=0$ give rise to the singular part $\check{\mathcal{D}}^{\text{sing}}$ of the diffuson matrix at small $\bm{q},\omega$. To compute $\check{\mathcal{D}}^{\text{sing}}$, we calculate the dispersion $\lambda_{l}(\bm q,\omega)$ of these zero modes by using the gradient expansion of Eq.~\eqref{eq:KD}. 

Assuming that $\lambda_{l>0}\neq 0$, $\lambda_{0}(\bm q,\omega)$ can be derived via the non-degenerate perturbation theory. We first find that
\begin{equation}\label{eq:Kphi}
\begin{split}
    (\check{K}^{D}(\bm{q},\omega)\vec{\psi}^{0})_{\bm{k} i j} = &\sum_{\bm{p}i'j'}W_{\bm{k}ij_{1},\bm{p}i'j'}\left\{\omega \hat{G}^{R}(\hat{G}^{R}-\hat{G}^{A}) - \frac{1}{2}\sum_{\alpha}q_{\alpha}(\hat{G}^{R}\hat{v}^{\alpha}\hat{G}^{R}-2\hat{G}^{R}\hat{v}^{\alpha}\hat{G}^{A}+\hat{G}^{A}\hat{v}^{\alpha}\hat{G}^{A})\right. \\
    & \left. - \frac{1}{4}\sum_{\alpha\beta}q_{\alpha}q_{\beta}\left(\hat{G}^{R}\hat{v}^{\alpha}\hat{G}^{R}\hat{v}^{\beta}\hat{G}^{R}-\hat{G}^{R}\hat{v}^{\alpha}\hat{G}^{R}\hat{v}^{\beta}\hat{G}^{A}+\hat{G}^{R}\hat{v}^{\alpha}\hat{G}^{A}\hat{v}^{\beta}\hat{G}^{A}-\hat{G}^{A}\hat{v}^{\alpha}\hat{G}^{A}\hat{v}^{\beta}\hat{G}^{A}\right)\right\}_{\bm{p}i'j'}.
\end{split}
\end{equation}
In this equation, $(\hat{A}\hat{B})_{\bm{k}ij}\equiv (\hat{A}(\bm k)\hat{B}(\bm k))_{ij}$ and the frequency variable $\epsilon$ is not written explicitly to simplify the notation. We made an approximation $-\partial_{\bm p}(\hat{G}^{R/A})^{-1} =\partial_{\bm p} \hat{H}^{0} + \partial_{\bm p} \hat{\Sigma}^{R/A} \approx \hat{\bm v}(\bm p)$ as the self-energies vary much more slowly than the band dispersion in the momentum space. We can further drop terms that do not contain $\hat{G}^{R}$ and $\hat{G}^{A}$ simultaneously. To understand this, notice that in Eq.~\eqref{eq:det} the poles of $\hat{G}^{R}(\bm{p},\epsilon)$ near a point $\bm{p}_0$ on the Fermi surface are distributed on one side of the real axis in the complex plane of $\xi_{n\bm{p}}$ because $\Gamma_{n\bm{p}}>0, \forall\{n,\bm{p}\}$. Thus those dropped terms are suppressed by a factor of $\mathcal{O}(\Gamma_{n\bm{p}}/\epsilon_{F})$ after integrating over $\bm p$ around $\bm{p}_0$. 

By applying Eq.~\eqref{eq:self2}, we arrive at
\begin{align}\label{eq:K1_1}
    \cev{\psi}^{0}\check{K}^{D}(\bm{q},\omega)\vec{\psi}^{0} = -\frac{2\pi i\gamma\Omega\omega}{\mathcal{N}} + \frac{1}{\mathcal{N}}\sum_{\bm k}\sum_{\alpha\beta}q_\alpha q_\beta\text{tr}\left(\hat{v}^{\alpha}(\bm k)\hat{G}^{R}(\bm k)\hat{v}^{\beta}(\bm k)\hat{G}^{A}(\bm k)\right) + ...,
\end{align}
Note that the linear-in-$q$ term in the expansion of $\check{K}^{D}(\bm q,\omega)$ does not contribute to Eq.~\eqref{eq:K1_1} due to the time-reversal symmetry, but it may couple the $l=0$ mode with high-frequency modes ($l> 0$),
\begin{align}
    &\cev{\psi}^{l}\check{K}^{D}(\bm{q},\omega)\vec{\psi}^{0} = \frac{1}{\mathcal{N}}\sum_{\alpha}q_{\alpha}\sum_{\bm{k}ij,\bm{k}'i'j'}\cev{\psi}_{\bm{k}ij}^{l}W_{\bm{k}ij,\bm{k}'i'j'}\left(\hat{G}^{R}(\bm{k}')\hat{v}^{\alpha}(\bm{k}')\hat{G}^{A}(\bm{k}')\right)_{i'j'},\label{eq:Kl0}\\
    &\cev{\psi}^{0}\check{K}^{D}(\bm{q},\omega)\vec{\psi}^{l} = \frac{1}{\mathcal{N}}\sum_{\alpha}q_{\alpha}\sum_{\bm{k}ij}\left(\hat{G}^{A}(\bm{k})\hat{v}^{\alpha}(\bm{k})\hat{G}^{R}(\bm{k})\right)_{ji}\vec{\psi}_{\bm{k}ij}^{l},\label{eq:K0l}
\end{align}
This coupling generates a quardratic-in-$\bm q$ correction to $\lambda_0(\bm q,\omega)$ through the second-order perturbation,
\begin{equation}
    \sum_{l\neq 0}\cev{\psi}^{0}\check{K}^{D}(\bm{q},\omega)\vec{\psi}^{l}\frac{1}{-\lambda_{l}}\cev{\psi}^{l}\check{K}^{D}(\bm{q},\omega)\vec{\psi}^{0} = \frac{1}{\mathcal{N}}\sum_{\alpha\beta}q_{\alpha}q_{\beta}\left[(\hat{G}^{A}\hat{v}^{\alpha}\hat{G}^{R})\cdot\check{\Pi}\frac{1}{\check{K}^{D}(0,0)}\check{\Pi}(\mathbf{1}-\check{K}^{D}(0,0))\cdot\vec{v}^{\beta}\right]. \label{eq:K1_2}
\end{equation}
Here we defined $(\vec{v}^{\beta})_{\bm{k} ij}\equiv v_{ij}^{\beta}(\bm{k})$. The projection operator $\check{\Pi}$ projects out the zero eigenspace of $\check{K}(0,0)$. Importantly, $\vec{v}^{\beta}$ is outside the zero eigenspace, $\check{\Pi}\cdot\vec{v}^{\beta}=\vec{v}^{\beta}$, because the time-reversal symmetry ensures $\cev{\psi}_{\bm{k}ij}^{0}\cdot\vec{v}^{\beta}=0$. Adding up Eqs.~\eqref{eq:K1_1} and \eqref{eq:K1_2} and simplifying the result with the help of $\vec{\Lambda}^{\beta}= \check{\Pi}\check{K}^{D}(0,0)^{-1}\check{\Pi}\cdot\vec{v}^{\beta}$ derived from Eq.~\eqref{eq:vertex_v}, we arrive at
\begin{equation}\label{eq:lambda}
    \lambda_{0}(\bm{q},\omega)=\frac{2\pi\gamma}{\mathcal{N} }\Big(-i\omega+\sum_{\alpha\beta}D^{\alpha\beta}q_\alpha q_\beta\Big),
\end{equation}
with the symmetric diffusion tensor $D^{\alpha\beta}=\sum_{\bm{k}}\text{tr}\left[\hat{v}^{\alpha}(\bm{k})\hat{G}^{R}(\bm{k})\hat{\Lambda}^{\beta}(\bm{k})\hat{G}^{A}(\bm{k})\right]/2\pi\gamma\Omega=\sigma_{D}^{\alpha\beta}/e^2\gamma$. The diffuson matrix at small $\omega$ and $q$ reads that
\begin{equation} \label{eq:diffuson_singular}
    \mathcal{D}_{\bm k i j, \bm{k}'i'j'}^{\text{sing}}(\bm q ,\omega)=\frac{1}{2\pi\gamma}\frac{\Delta\Sigma_{ij}(\bm k)\Delta\Sigma_{i'j'}(\bm{k}')^{*}}{-i\omega+\sum\limits_{\alpha\beta}D^{\alpha\beta}q_{\alpha} q_{\beta}}.
\end{equation}

\subsection{Cooperons}

Similar to the diffuson, Cooperon ladder sum in Fig.~\ref{fig:diffuson}a satisfies the integral equation $\check{K}(\bm q,\omega)\check{C}=\check{U}$, where $U_{\bm{k}ij,\bm{k'}i'j'}(\bm q)\equiv\bra{\bm{k}_{+}i,-\bm{k}_{-}j}U\ket{\bm{k}_{+}' i',-\bm{k}_{-}'j'}$. When the $\bm q$-dependence of $\check{U}$ is neglected,
the kernel $\check{K}$ reads that
\begin{equation}
    K_{\bm{k}ij,\bm{k}'i'j'}(\bm q,\omega)=\delta_{\bm k,\bm{k}'}\delta_{i,i'}\delta_{j,j'}-\sum_{i_1j_1}U_{\bm{k}ij,\bm{k}'i_{1}j_{1}}G_{i_1,i'}^{R}\left(\bm{k}_{+}',\epsilon+\omega\right)G_{j_1,j'}^{A}\left(-\bm{k}_{-}',\epsilon\right). \label{eq:kernel}
\end{equation}
A generalized time-reversal symmetry $\mathcal{T}=U_{T}\mathcal{K}$ relates the Cooperon to the diffuson via a unitary transformation,
\begin{align}
    C_{\bm{k}ij,\bm{k}'i'j'}(\bm q,\omega) &= \sum_{j_1,j_1'}(U_{T})_{jj_1}\mathcal{D}_{\bm{k}ij_{1},\bm{k}'i'j_1'}(\bm q,\omega)(U_{T}^{\dagger})_{j_1'j'},\label{eq:dtoc}\\
    K_{\bm{k}ij,\bm{k}'i'j'}(\bm q,\omega) &= \sum_{j_1,j_1'}(U_{T})_{jj_1}K_{\bm{k}ij_{1},\bm{k}'i'j_1'}^{D}(\bm q,\omega)(U_{T}^{\dagger})_{j_1'j'}.
\end{align}
Therefore, $\check{K}$ and $\check{K}^{D}$ share the same eigenvalues. The right(left) eigenfunctions $\vec{\phi}(\cev{\phi})$ of $\check{K}(0,0)$ can be derived by a unitary transformation upon $\vec{\psi}(\cev{\psi})$,  $\vec{\phi}_{\bm{k} i j}^{l}=\sum_{j'}(U_{T})_{jj'}\vec{\psi}_{\bm{k}ij'}^{l}$ and $\cev{\phi}_{\bm{k} i j}^{l}=\sum_{j'}\cev{\psi}_{\bm{k}ij'}^{l}(U_{T}^{\dagger})_{j'j}$.
In particular, $\check{K}(0,0)$ has a zero eigenvalue associated with a right eigenfunction 
\begin{equation}\label{eq:phi0}
    \vec{\phi}_{\bm{k}ij}^{0}=\left(\Delta\hat{\Sigma}(\bm k)U_{T}^{\text{t}}\right)_{ij}=\left(U_{T}\Delta\hat{\Sigma}(\bm k)^{*}\right)_{ji},
\end{equation}
and the left eigenfunction is $\cev{\phi}_{\bm{k}ij}^{0}=\frac{1}{\mathcal{N}}\left(\Delta\hat{G}(\bm{k}')^{\text{t}}U_{T}^{\dagger}\right)_{ij}$. Plugging Eqs.~\eqref{eq:diffuson} and~\eqref{eq:diffuson_singular} into Eq.~\eqref{eq:dtoc}, we obtain the Cooperon matrix at small $\omega$ and $q$ 
\begin{equation}\label{eq:cooperon_decomp}
    C_{\bm k i j, \bm{k}'i'j'}(\bm q ,\omega)\approx\sum_{\substack{l\\ \lambda_l\neq 0}}\frac{1}{\lambda_l}\vec{\phi}_{\bm k i j}^{l}\tilde{\phi}_{\bm k' i' j'}^{l} + C_{\bm k i j, \bm{k}'i'j'}^{\text{sing}}(\bm q ,\omega),
\end{equation}
with $\tilde{\phi}_{\bm k i j}^{l}=\sum_{\bm p i_{1}j_{1}}\cev{\phi}_{\bm p i_{1}j_{1}}^{l}U_{\bm p i_{1}j_{1}, \bm k i j}$ and the singular part of the Cooperon matrix
\begin{equation}\label{eq:cooperon_singular} 
    C_{\bm k i j, \bm{k}'i'j'}^{\text{sing}}(\bm q ,\omega)=\frac{1}{2\pi\gamma}\frac{\left(\Delta\hat{\Sigma}(\bm k)U_{T}^{\text{t}}\right)_{ij}\left(\Delta\hat{\Sigma}(\bm{k}')U_{T}^{\text{t}}\right)_{i'j'}^{*}}{-i\omega+\sum\limits_{\alpha\beta}D^{\alpha\beta}q_{\alpha} q_{\beta}}.
\end{equation}
Owing to time-reversal symmetry,
\begin{equation}\label{eq:cooperon_trs}
    C_{-\bm{k}ij,-\bm{k}'i'j'}(-\bm{q},\omega)=\sum_{i_1,j_1,i_2,j_2}(U_{T})_{ii_1}(U_{T})_{jj_1}C_{\bm{k}'i_2j_2,\bm{k}i_1j_1}(\bm{q},\omega)(U_{T}^{\dagger})_{i_2i'}(U_{T}^{\dagger})_{j_2j'}.
\end{equation}
Plugging Eq.~\eqref{eq:cooperon_decomp} into the above equation reveals the relation $\tilde{\phi}_{-\bm{k}ij}^{l}\propto\sum_{i_1j_1}(U_{T}^{*})_{ii_1}\vec{\phi}_{\bm{k} i_{1}j_{1}}^{l}(U_{T}^{\dagger})_{j_1j}$.

\subsection{Quantum corrections to the conductivity}
\begin{figure}
    \centering
    \includegraphics[width=0.75\linewidth]{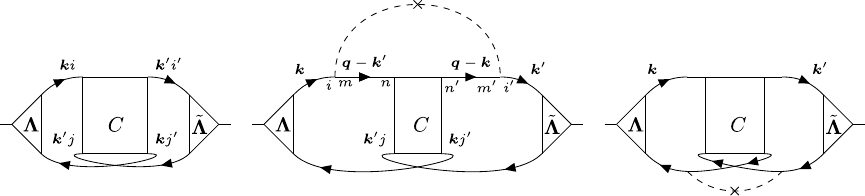}
    \caption{Leading contributions to the scale-dependent conductance corrections in the weak-scattering limit. The first diagram is the bare ``Hikami box" and the second and third diagrams are dressed ``Hikami boxes".}
    \label{fig:hikami_box}
\end{figure}
The WL correction to dc conductance equals the sum of three ``Hikami" boxes in Fig.~\ref{fig:hikami_box} \cite{mccann2006weak},
\begin{align}
    \delta \sigma^{\alpha\beta}=& e^2\int \frac{d \epsilon}{2 \pi} \frac{\partial n_{F}}{\partial \epsilon}\sum_{\bm k ij, \bm{k}' i'j^{\prime}} \left(\hat{G}^{A}(\bm{k}')\tilde{\Lambda}^{\alpha}(\bm{k}') \hat{G}^{R}(\bm{k}')\right)_{ji'} \left(\hat{G}^{R}(\bm{k}) \hat{\Lambda}^{\beta}(\bm{k}) \hat{G}^{A}(\bm{k})\right)_{ij'}\left[C_{\frac{\bm k-\bm{k}'}{2} i'j',\frac{\bm k'-\bm k}{2} ij}(\bm k+\bm{k}',0)\right.\notag\\
    &\qquad+\sum_{\bm q, mn,m'n'}G_{m'n'}^{R}(\bm q-\bm k)C_{\frac{\bm q}{2}-\bm k n'j',\frac{\bm q}{2}-\bm{k}' nj}(\bm q,0)G_{nm}^{R}(\bm q-\bm{k}')\bra{\bm{q}-\bm{k}'m,\bm{k}'i'}U\ket{\bm{k}i,\bm{q}-\bm{k}m'}+c.c.\Big], \label{eq:deltasigma}
\end{align}
Assuming that $\check{K}$ has a single gapless eigenmode $\lambda_{0}(\bm q,\omega)$, we can replace the Cooperon matrix by its singular part $\check{C}^{\textrm{sing}}(\bm{q},0)$, Eq.~\eqref{eq:cooperon_singular}, in the long wavelength limit $q<q_c$ where $\lambda_0(\bm q_c,0)\sim \min_{l>0}|\lambda_l|$. By setting $\bm{q}=0$ in Eq.~\eqref{eq:deltasigma} except for $\check{C}^{\textrm{sing}}(\bm{q},0)$, we simplify the equation to
\begin{equation}\label{eq:deltasigma_n}
    \delta \sigma^{\alpha\beta}\approx\frac{e^2}{\hbar}\int \frac{d \epsilon}{2 \pi} \frac{\partial n_{F}}{\partial \epsilon}\left(N_{1}^{\alpha\beta}(\epsilon)+N_{2}^{\alpha\beta}(\epsilon)+N_{2}^{\alpha\beta}(\epsilon)^{*}\right)\int \frac{d^{2} q}{(2\pi)^{2}} \frac{1}{\bm q\cdot D\cdot\bm q + \tau_{\phi}^{-1}}+....
\end{equation}
where `...' contain non-singular conductance corrections. $N_1 (N_2)$ corresponds to the frist (second) term in the square bracket in Eq.~\eqref{eq:deltasigma},
\begin{align}
    N_{1}^{\alpha\beta}=&\frac{1}{2\pi\gamma\Omega}\sum_{\bm k} \operatorname{tr}\left\{ \hat{G}^{A}(\bm k)\tilde{\Lambda}^{\alpha}(\bm{k}) \hat{G}^{R}(\bm k) \Delta\hat{\Sigma}(\bm k)U_{T}^{\text{t}}\left[\hat{G}^{R}(-\bm k) \hat{\Lambda}^{\beta}(-\bm{k}) \hat{G}^{A}(-\bm k)\right]^{\text{t}}\Delta\hat{\Sigma}(-\bm k)^{*}U_{T}^{\dagger}\right\}\notag\\
    =&\frac{1}{2\pi\gamma\Omega}\sum_{\bm k} \operatorname{tr}\left\{\hat{G}^{A}(\bm k)\tilde{\Lambda}^{\alpha}(\bm{k}) \hat{G}^{R}(\bm k) \Delta\hat{\Sigma}(\bm k)\hat{G}^{A}(\bm k) \hat{\Lambda}^{\beta}(\bm{k}) \hat{G}^{R}(\bm k)\Delta\hat{\Sigma}(\bm k) U_{T}^{\text{t}}U_{T}^{\dagger}\right\}\notag\\
    =&\frac{s}{2\pi\gamma\Omega}\sum_{\bm k}\operatorname{tr}\left[\tilde{\Lambda}^{\alpha}(\bm k)\left(\hat{G}^{R}(\bm k)-\hat{G}^{A}(\bm k)\right)\hat{\Lambda}^{\beta}(\bm k)\left(\hat{G}^{R}(\bm k)-\hat{G}^{A}(\bm k)\right)\right]\notag\\
    \approx&-\frac{s}{2\pi\gamma\Omega}\sum_{\bm k}\operatorname{tr}\left[\tilde{\Lambda}^{\alpha}(\bm k)\hat{G}^{R}(\bm k)\hat{\Lambda}^{\beta}(\bm k)\hat{G}^{A}(\bm k) + \tilde{\Lambda}^{\alpha}(\bm k)\hat{G}^{A}(\bm k)\hat{\Lambda}^{\beta}(\bm k)\hat{G}^{R}(\bm k)\right], \label{eq:n1}
\end{align}
\begin{align}
    N_{2}^{\alpha\beta}=& \frac{1}{2\pi\gamma\Omega^2}\sum_{\bm k\bm k', mm',ii'}\bra{-\bm{k}'m,\bm{k}'i'}U\ket{\bm{k}i,-\bm{k}m'} \left[\hat{G}^{R}(\bm k)\hat{\Lambda}^{\beta}(\bm{k}) \hat{G}^{A}(\bm k)\left(\hat{G}^{R}(-\bm k)\Delta\hat{\Sigma}(-\bm k)U_{T}^{\text{t}}\right)^{\textrm{t}}\right]_{im'}\notag\\
    &\qquad\qquad\quad \times\left[\left(\Delta\hat{\Sigma}(-\bm{k}')^{*}U_{T}^{\dagger}\hat{G}^{A}(\bm{k}') \tilde{\Lambda}^{\alpha}(\bm{k}') \hat{G}^{R}(\bm{k}')\right)^{\textrm{t}}\hat{G}^{R}(-\bm{k}')\right]_{i'm} \notag\\
    \approx &\frac{s}{2\pi\gamma\Omega^2}\sum_{\bm k\bm{k}', mm',ii'} \left(\hat{G}^{R}(-\bm{k}')\tilde{\Lambda}^{\alpha}(-\bm{k}')\hat{G}^{A}(-\bm{k}')\right)_{i'm}W_{-\bm{k}' mi',\bm k im'}\left(\hat{G}^{R}(\bm k)\hat{\Lambda}^{\beta}(\bm k)\hat{G}^{A}(\bm k)\right)_{im'}\notag\\
    =&\frac{s}{2\pi\gamma\Omega}\sum_{\bm k}\operatorname{tr}\left[\left(\tilde{\Lambda}^{\alpha}(\bm k)-\hat{v}^{\alpha}(\bm k)\right)\hat{G}^{R}(\bm k)\hat{\Lambda}^{\beta}(\bm k)\hat{G}^{A}(\bm k)\right], \label{eq:n2_1}\\
    =&\frac{s}{2\pi\gamma\Omega}\sum_{\bm k}\operatorname{tr}\left[\tilde{\Lambda}^{\alpha}(\bm k)\hat{G}^{A}(\bm k)\left(\hat{\Lambda}^{\beta}(\bm k)-\hat{v}^{\beta}(\bm k)\right)\hat{G}^{R}(\bm k)\right], \label{eq:n2_2}
\end{align}
with $s=\mathcal{T}^2$. As explained before, we dropped the terms that contain only $\hat{G}^{R}$ or $\hat{G}^{A}$ but not both. We applied the self-consistent equations for $\hat{\bm\Lambda}$ and $\tilde{\bm\Lambda}$, Eqs.~\eqref{eq:vertex_v} and~\eqref{eq:vertex_vtilde}, to derive Eqs.~\eqref{eq:n2_1} and~\eqref{eq:n2_2}, respectively. $N_2=N_2^{*}$ due to the hermiticity of $\hat{\bm\Lambda}$ and $\tilde{\bm\Lambda}$. 

Plugging Eqs.~\eqref{eq:n1}-\eqref{eq:n2_2} into Eq.~\eqref{eq:deltasigma} and then using Eqs.~\eqref{eq:sigmad} and~\eqref{eq:sigmad_2} for the Drude conductivity, we obtain the main result of this section, the interference conductance correction,
\begin{equation}\label{eq:deltasigma_2}
    \delta\sigma^{\alpha\beta}=-s\frac{2e^2}{h}D^{\alpha\beta}\int_{|\bm q|<q_{c}} \frac{d^{2} q}{(2\pi)^2} \frac{1}{\bm q\cdot D\cdot\bm q+ \tau_{\phi}^{-1}} + ....
\end{equation}
This equation reveals that the conductance corrections along arbitrary in-plane axes have an infrared logarithmic divergence in 2D (cut off by the decoherence $\tau_{\phi}^{-1}$), whose sign is entirely determined by $\mathcal{T}^2$ since $D^{\alpha\beta}=\sigma^{\alpha\beta}/\gamma$ is positive definite. 

In a weak out-of-plane magnetic field $\bm{B}=\bm{\nabla}\cross\bm{A}$, $\bm q$ is replaced by $-i\bm{\nabla}+2e\bm{A}$ and the eigenvalue $\lambda_{0}(\bm q,\omega)$ is quantized into Landau levels, $4eB(n+1/2)\sqrt{\det D}/\hbar,\ n\in \mathbb{N}$. Thus, magnetic field can generate a finite Cooperon gap, reduce the magnitude of interference corrections, and yield a positive (negative) magnetoconductance $\Delta\sigma(B)=\delta\sigma(B)-\delta\sigma(0)$ in time-reversal invariant systems with $\mathcal{T}^2=1 (-1)$. For $B\ll \hbar/2eq_{c}^2$,
\begin{align}
    \Delta\sigma^{\alpha\beta}(B) &= -s\frac{2e^2}{h}D^{\alpha\beta}\left\{\operatorname{Tr}\left[\frac{1}{(-i\bm{\nabla}+2e\bm{A})\cdot\bm{D}\cdot(-i\bm{\nabla}+2e\bm{A})+\tau_{\phi}^{-1}}\right]-\operatorname{Tr}\left[\frac{1}{-\bm{\nabla}\cdot\bm{D}\cdot\bm{\nabla}+\tau_{\phi}^{-1}}\right]\right\}\notag\\
    &=\frac{se^2}{2\pi h}\frac{\sigma_{D}^{\alpha\beta}}{\sqrt{\det \sigma_{D}}}F\left(\frac{B}{B_{\phi}}\right),\qquad F(x)=\ln{x}+\psi\left(\frac{1}{2}+\frac{1}{x}\right). \label{eq:magneto}
\end{align}
Here, $\psi$ is the digamma function, and $B_{\phi}\equiv \hbar/4e\tau_{\phi}\sqrt{\det D}$ determines the curvature of $\Delta\sigma$ at $B\lesssim B_{\phi}$.

\section{Generalization to graphene multilayers}
%
%

The interference conductance correction Eq.~\eqref{eq:deltasigma} and the weak-field magnetoresistance Eq.~\eqref{eq:magneto} indicate that WL and WAL effects are contingent on the time-reversal symmetry and are universal phenomena in time-reversal invariant diffusive metallic systems. However, our derivation only considers a single gapless Cooperon mode and needs to be generalized to systems with more conserved quantum numbers than the particle number, such as spin and valley numbers. Graphene has an approximate spin SU(2) symmetry due to the negligible spin-orbit coupling, so each spin flavor retains the orbital time-reversal symmetry. Since different spin channels conduct in parallel, the total conductance correction equals the correction in a single spin channel multiplied by the spin degeneracy $g_s$. In this section, we therefore focus on the valley degrees of freedom in a single spin. Additionally, we include less singular but experimentally observable contributions from weakly gapped Cooperons into the magnetoconductance formula to elucidate the rise of WL or WAL effects as the intervalley coherent (IVC) order gradually develops.
\\
\\
\noindent\textbf{Valley-number conserved normal states: }
Because trigonal warping of energy bands breaks the valley SU(2) symmetry, individual valley does not enjoy any generalized time-reversal symmetries. Instead, the ordinary time-reversal $\mathcal{T}_O (\mathcal{T}_O^2=1)$ transforms one valley to the other and two valleys together can host gapless Cooperon modes. The combination of $\mathcal{T}_{O}$ and the valley U(1) symmetry $\tau^z$ gives rise to the Kramers time-reversal symmetry $\mathcal{T}_{K}=-i\tau^z\mathcal{T}_{O}$ with $\mathcal{T}_{K}^2=(\tau^z\mathcal{T}_{O})^2=-1$. According to Eq.~\eqref{eq:phi0}, the kernel has two-fold degenerate zero eigenvalues associated with eigenfunctions $\vec{\phi}_{\bm{k}ij}^{l}=(\Delta\hat{\Sigma}(\bm k) U_{T,l}^{\text{t}})_{ij}\ (l=0,1,\text{and } U_{T,1}=\tau^{z}U_{T,0})$. These two modes are orthogonal,
\begin{equation}
    \cev{\phi}^{0}\cdot\vec{\phi}^{1}=\frac{1}{\mathcal{N}\Omega}\sum_{\bm k}\tr[U_{T,0}^{*}\Delta\hat{G}(\bm k)\Delta\hat{\Sigma}(\bm k)U_{T,1}^{\text{t}}]=-\frac{1}{\mathcal{N}\Omega}\sum_{\bm k}\tr[\Delta\hat{G}(-\bm k)^{\text{t}}U_{T,0}U_{T,1}^{\dagger}\Delta\hat{\Sigma}(-\bm k)^{\text{t}}] = -\cev{\phi}^{0}\cdot\vec{\phi}^{1} =0.
\end{equation}
In the second equation, we used both time-reversal symmetries $U_{T,0}\Delta G(\bm{k})^{\text{t}}U_{T,0}^{\dagger}=\Delta G(-\bm{k})$ and $U_{T,1}\Delta\Sigma(\bm{k})^{\text{t}}U_{T,1}^{\dagger}=\Delta\Sigma(-\bm{k})$ with $U_{T,0}^{*}U_{T,0}=-U_{T,1}^{*}U_{T,1}=1$. By the same methods, we can prove that all terms of even orders in $\bm q$ in the gradient expansion of $\cev{\phi}^{0}\check{K}(\bm{q},\omega)\vec{\phi}^{1}$ must vanish. Owing to the relation $\cev{\phi}^{0}\check{K}(\bm{q},\omega)\vec{\phi}^{1}=\cev{\psi}^{0}\check{K}^{D}(\bm{q},\omega)\vec{\psi}^{1}$, we can deduce from Eq.~\eqref{eq:K0l}
\begin{align}
    \cev{\phi}^{0}\check{K}(\bm{q},\omega)\vec{\phi}^{1}&=\frac{1}{\mathcal{N}\Omega}\sum_{\bm k}\bm q\cdot\tr\left(U_{T,0}^{\text{t}}\hat{G}^{A}(\bm{k})\hat{\bm v}(\bm k)\hat{G}^{R}(\bm{k})\Delta\hat{\Sigma}(\bm{k}) U_{T,1}^{*}\right)+O(q^3)\notag\\
    &=\frac{1}{\mathcal{N}\Omega}\sum_{\bm k}\bm q\cdot\tr\left[i\tau^{z}\hat{\bm v}(\bm k)\left(\hat{G}^{A}(\bm{k})-\hat{G}^{R}(\bm{k})\right)\right]+O(q^3)\notag\\
    &=\frac{1}{\mathcal{N}\Omega}\sum_{\bm k}\bm q\cdot\nabla_{\bm k}\tr\left[i\tau^{z}\left(\ln\hat{G}^{R}(\bm{k})-\ln\hat{G}^{A}(\bm{k})\right)\right]+O(q^3)=O(q^3).
\end{align}
We see that the linear-in-$\bm q$ term also vanishes. In conclusion, valley symmetric states have two decoupled gapless Cooperon modes $\vec{\phi}^{0,1}$ which contribute oppositely to the conductance because $T_{O}^2=-\mathcal{T}_{K}^2=1$. Graphene will not exhibit WL or WAL if the electron valley number is conserved.
\\
\\
\noindent\textbf{Intervalley coherent states: }
When the valley-number conservation is broken by time-reversal invariant IVC order, either $\mathcal{T}_{O}$ or $\mathcal{T}_{K}$ symmetry must be violated. Without loss of generality, let us consider that $\mathcal{T}_{K}$ symmetry is weakly broken but $\mathcal{T}_{O}$ preserved. The Cooperon gaps correspondingly become $\lambda_1 \neq \lambda_0=0$. Note that $\lambda_1\in \mathbb{R}$ for weak IVC order, otherwise complex eigenvalues $\lambda$ must come in conjugate pairs (see \ref{prop2}) but by continuity we expect to see a single weakly gapped Cooperon mode, \textit{i.e.}, $0=\lambda_0<\lambda_1\ll |\lambda_{l\neq 0,1}|$. Eq.~\eqref{eq:cooperon_trs} constrains the possible forms of the Cooperon matrix projected into the two-dimensional space spanned by $\phi_{\bm{k}ij}^{0,1}$,
\begin{equation}\label{eq:cooperonv}
    C_{\bm k i j, \bm{k}'i'j'}^{\text{sing}}(\bm q ,0)=\frac{\mathcal{N}}{2\pi\gamma}\left(\vec{\phi}_{\bm{k} i j}^{0},\vec{\phi}_{\bm{k} i j}^{1}\right)
    \begin{pmatrix}
    \bm{q}\cdot D_{00}\cdot\bm{q} + \tau_{\phi}^{-1} & i\bm{c}\cdot \bm{q} + \bm{q}\cdot  D_{01}\cdot \bm{q} \\
    -i\bm{c}\cdot \bm{q} + \bm{q}\cdot  D_{01}\cdot \bm{q} & \bm{q}\cdot D_{11}\cdot\bm{q} + \frac{\mathcal{N}\lambda_1}{2\pi\gamma} + \tilde{\tau}_{\phi}^{-1}
    \end{pmatrix}^{-1}
    \begin{pmatrix}
    \tilde{\phi}_{\bm{k}' i' j'}^{0}\\
    \tilde{\phi}_{\bm{k}' i' j'}^{1}
    \end{pmatrix},
\end{equation}
where $\bm{c}$ is a real vector and $D_{ll^{'}}$'s are real two-dimensional tensors. 
According to our analysis above, $\bm{c}=0$ and $D_{01}=0$ without IVC order, and are therefore expected to remain small for weak IVC order. Besides, $\bm{c}=0$ in $C_{3z}$ symmetric systems because one can show that the eigenfunctions $\vec{\phi}^{0,1}(\cev{\phi}^{0,1})$ of the $C_{3z}-$invariant kernel $\check{K}(0,0)$ are also $C_{3z}-$invariant, $\sum_{i'j'}(C_{3z})_{ii'}(C_{3z})_{jj'}\vec{\phi}_{\bm{k}i'j'}^{0,1} = \vec{\phi}_{C_{3z}\bm{k}ij}^{0,1}$ \footnote{The $C_{3z}$-invariance of $\vec{\phi}^{0}$ can be proven using Eq.~\eqref{eq:phi0} and the commutation relation between $\mathcal{T}$ and $C_{3z}$, $U_{T}C_{3z}^{*}=C_{3z}U_{T}$. By the same method, we can verify the invariance of $\vec{\phi}^{1}$ in valley symmetric phase. Because quantized $C_{3z}$ charge cannot be changed by weak IVC order, $\vec{\phi}^{1}$ remain $C_{3z}$ invariant when the IVC order parameter is sufficiently small.}, and therefore Eqs.~\eqref{eq:Kl0} and \eqref{eq:K0l} must vanish. 

When both $\bm{c}$ and $D_{01}$ are neglected, $\vec{\phi}^{0}$ and $\vec{\phi}^{1}$ are decoupled and $D_{00}=D$, leading to the following weak-field magnetoconductance formula,
\begin{equation}\label{eq:deltasigma_B}
    \Delta\sigma^{\alpha\beta}(B)=\frac{se^2}{2\pi h}\frac{\sigma_{D}^{\alpha\beta}}{\sqrt{\det \sigma_{D}}}\left(F\left(\frac{B}{B_{\phi}}\right)-\Xi F\left(\frac{B}{B_{\phi}+B_{v}}\right)\right).
\end{equation}
In the valley-number conserved normal states, $\Xi=1$, $B_{v}=0$, and $\Delta\sigma^{\alpha\beta}(B)=0$. For generic IVC states, $\Xi$ can in principle deviate from 1 and become nonuniversal, although there are examples with $\Xi= 1$ (see below). $B_{v} = (\mathcal{N}\lambda_1/2\pi\gamma+\tilde{\tau}_{\phi}^{-1}-\tau_{\phi}^{-1})/4e\sqrt{\det D_{11}}  \approx \mathcal{N}\lambda_1/8\pi\gamma e\sqrt{\det D_{11}}$. The last approximation is based on the observation that $\tau_\phi=\tilde{\tau}_\phi$ in the normal state \cite{mccann2006weak} and $\tau_\phi^{-1},\tilde{\tau}_\phi^{-1}\ll \frac{\mathcal{N}\lambda_1}{2\pi\gamma}$ when IVC order becomes strong.

In experiment, one can fit the low-temperature weak-field magnetoconductance data with Eq.~\eqref{eq:deltasigma_B} to determine the three fitting parameters $B_{\phi}, B_{v},\Xi$. If the WL effect is generated by intervalley scattering potential \cite{mccann2006weak}, $\Xi=1$, and $L_v=\sqrt{\hbar/4eB_v}$ should be comparable and sensitive to the sample dimensions \cite{tikhonenko2008weak}. In contrast, if the WL effect is enabled by IVC order, $\lambda_1\rightarrow 1$ and $L_v$ could become as short as the mean-free path.  

\subsection{ABC trilayer graphene in the strong displacement field}

\begin{figure}
    \centering
    \includegraphics[width=0.6\linewidth]{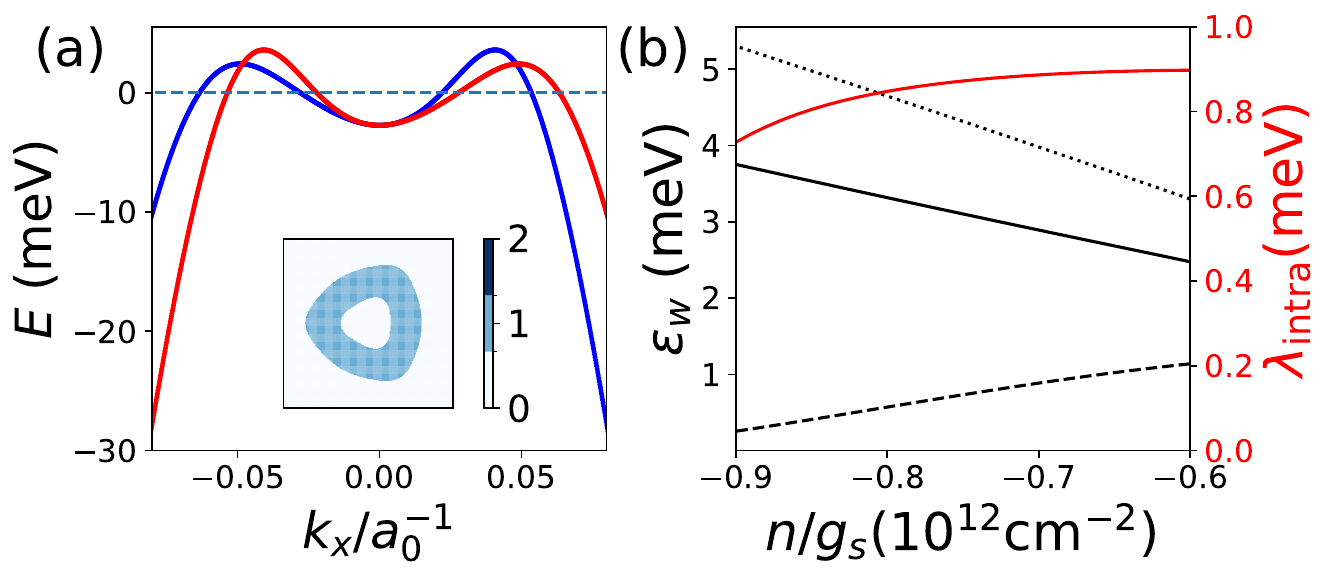}
    \caption{(a) The electronic structure of a normal metallic state in ABC trilayer graphene with potential difference between two outer layers $U=40$\si{meV}. The blue(red) line is the $K(-K)-$valley diserpsion $\epsilon_{\bm{k},+}(\epsilon_{\bm{k},-})$ and the dashed line is the Fermi level at hole density per spin $-n/g_s=7\times10^{11}$\si{\centi\meter^{-2}}. The inset plots the $K-$valley hole occupation number in the momentum space and shows an annulus Fermi surface. (b) The black lines characterize the average trigoanl warping energies along different Fermi surfaces, $\epsilon_w = (\sum_{\bm k\in \textrm{F.S.}}\delta(\epsilon-\epsilon_{\bm k +})(\epsilon_{\bm{k},+}-\epsilon_{-\bm{k},+})^2)^{1/2}/(\gamma/2)$, as a function of carrier density per spin, where $\textrm{F.S.}$ stands for the inner, outer, and both Fermi surfaces for the dashed, dotted, and solid lines, respectively, and $\gamma$ is the total density of states per spin. The inner Fermi surface has relatively weak trigonal warping energy because of its closer distance to the Dirac point and larger layer polarization. The red line is the $\bm q=0$ intravalley Cooperon gap $\lambda_{\text{intra}}$.}
    \label{fig:trilayer_normal}
\end{figure}

In the main text, we study a simplified band Hamiltonian of ABC trilayer graphene in the strong displacement field , $\hat{H}^{0}(\bm k)=\bar{\epsilon}_{\bm k}\tau^{0}+\frac{1}{2}\bm \Delta_{\bm k}\cdot\bm\tau$, with $\tau^{0}$ and $\tau^{x,y,z}$ respectively representing the identity matrix and three Pauli matrices in the valley space, assuming that the layer and sublattice degrees of freedom of charge carriers are fully polarized by the strong displacement field. In the normal state, $\Delta_{\bm k}^{x,y}=0$. $\bar{\epsilon}_{\bm{k}}$ and $\Delta_{\bm{k}}^{z}$ are determined by single-particle tight-binding band structure illustrated in Fig.~\ref{fig:trilayer_normal}a. We take a short-range correlated random potential, $\bra{\bm{k}'i',\bm{k}+\bm{q}i}U\ket{\bm{k}'+\bm{q} j',\bm{k}j}=u_0^2\delta_{ij}$, where $i,j$ are valley indices. The disorder potentials is considered to vary smoothly over a length scale significantly longer than the graphene lattice constant, thereby preserving the valley degree of freedom of the electrons. We use Eq.~(9) in the main text to calculate the intravalley Cooperon gap $\lambda_{\textrm{intra}}$ as a function of hole density for $\hbar/\tau_{0}=0.2$\si{meV} ($\tau_0^{-1}\equiv\pi u_0^2\gamma$).  Fig.~\ref{fig:trilayer_normal}b shows that $\lambda_{\textrm{intra}}\approx 1$, indicating that the associated interference correction to conductance is suppressed by trigonal warping.

For the IVC phase, we use a constant valley XY-exchange potential $\Delta\tau^{x}$. The inner electron-like Fermi pocket has weaker trigonal warping energy (see Fig.~\ref{fig:trilayer_normal}b) and stronger intervalley mixing than the states on the outer Fermi surface (see Fig.~2a in the main text). Within the approximation that the valley XY-exchange potential $\Delta\gg \hbar/\tau_0$, the band splitting $|\bm\Delta_{\bm k}|$ is much larger than the disorder broadening for all $\bm k$ on Fermi surfaces and thus it is justified to neglect the disorder-induced band mixing and write down the following disorder averaged retarded and advanced Green's functions 
\begin{equation}\label{eq:GRA_appendix}
    \hat{G}^{R/A}(\bm k,\epsilon) = \sum_{l=\pm}\frac{|\bm{k}l\rangle\langle \bm{k}l|}{\epsilon-\epsilon_{\bm k l}\pm i\hbar/2\tau_{\bm k l}},
\end{equation}
Here, $\epsilon_{\bm k l}$ is the quasiparticle energy and $\tau_{\bm k l}$ is the relaxation time of the valley state $|\bm{k}l\rangle$ at momentum $\bm{k}$ in the energy band $l$ of $\hat{H}^{0}$. We find that there is a gapless Cooperon $\phi_{\bm{k}ij}^{0} = -i\sum_{\mu=0,x,y,z}\langle n^{\mu}\rangle (\tau^{\mu}\tau^{x})_{ij}/\tau_0$ for all $\Delta$ and a valley-singlet Cooepron $\phi_{\bm{k}ij}^{s} = \tau_{ij}^{y}/\tau_0$ with a gap $\lambda_s$ induced by IVC order. When these two Cooperons have the smallest gaps among the eigenmodes of the kernel $\check{K}$, which corresponds roughly to $\Delta\lesssim \epsilon_w$ of the inner Fermi surface (see the dashed line in Fig.~\ref{fig:trilayer_normal}b), the long-wavelength solution to the Bethe-Salpeter equation for the Cooperon reads that,
\begin{equation}\label{eq:cooperon_abc}
    C_{\bm k ij,\bm{k}'i'j'} (\bm{q},\omega=0)\approx \frac{1}{2\pi\gamma}\left(\frac{\phi_{\bm{k}ij}^{0}\phi_{\bm{k}'i'j'}^{0*}}{Dq^2}+\frac{\phi_{\bm{k}ij}^{s}\phi_{\bm{k}'i'j'}^{s*}}{D_sq^2+\lambda_s\tau_0^{-1}}\right)=\frac{1}{2\pi\gamma}\left(\frac{\phi_{\bm{k}ij}^{0}\phi_{\bm{k}'i'j'}^{0}}{Dq^2}-\frac{\phi_{\bm{k}ij}^{s}\phi_{\bm{k}'i'j'}^{s}}{D_sq^2+\lambda_s\tau_0^{-1}}\right),
\end{equation}
where $D_{s}=\sum_{\bm k}\sum_{l=\pm}\delta(\epsilon- \epsilon_{\bm{k}l})\bm{v}_{\bm{k}l}^2\tau_{\bm{k} l}/2\gamma\Omega$ and $D_{s}=\sum_{\bm k}\sum_{l=\pm}\delta(\epsilon- \epsilon_{\bm{k}l})\bm{v}_{\bm{k}l}^2 (n_{\bm{k}l}^{z})^2\tau_{\bm{k} l}^3/2\tau_0^2\gamma\Omega$. Here, $\gamma$ is the single-particle density of states per spin at energy $\epsilon$, $\bm v_{\bm{k}l} \equiv \langle \bm{k} l| \hat{v}(\bm k)|\bm{k} l\rangle$ is the group velocity and $n_{\bm{k} l}^{\mu}=\langle \bm{k} l|\tau^{\mu}|\bm{k} l\rangle$.

We now calculate the Feynman diagrams in Fig.~\ref{fig:hikami_box} to obtain the interference conductance correction. Plugging Eq.~\eqref{eq:GRA_appendix} into Eq.~\eqref{eq:vertex_v}, we notice that the disorder-induced current vertex correction vanishes in our model,
\begin{align}
    \hat{\bm\Lambda}_{\bm{k}} &=\hat{\bm{v}}_{\bm k} + \frac{u_0^2}{\Omega}\sum_{\bm{k}'}\sum_{l,l'=\pm}\frac{|\bm{k}'l\rangle\langle \bm{k}'l|}{\epsilon- \epsilon_{\bm{k}'l}+\frac{i}{2\tau_{\bm{k}'l}}}\hat{\bm\Lambda}_{\bm{k}'}\frac{|\bm{k}'l'\rangle\langle \bm{k}'l'|}{\epsilon- \epsilon_{\bm{k}'l'}-\frac{i}{2\tau_{\bm{k}'l'}}}\notag\\
    &=\hat{\bm{v}}_{\bm k} +\frac{2\pi u_0^2\tau}{\Omega}\sum_{\bm{k}'}\sum_{l=\pm}\delta(\epsilon-\epsilon_{\bm{k}'l})\langle \bm{k}'l| \hat{\bm\Lambda}_{\bm{k}'}|\bm{k}'l\rangle\frac{\sigma^{0}+\bm n_{\bm{k}'l}\cdot\bm\sigma}{2}\notag\\
    &=\hat{\bm{v}}_{\bm k}.
\end{align}
In the second line, we dropped the $l\neq l'$ cases due to the large band splitting. To arrive at the last equation, we noticed that the current vertex $\hat{\bm \Lambda}=(\hat{\Lambda}^{x},\hat{\Lambda}^{y})$ should be time-reversal odd and transform as a vector under $C_{3z}$ rotation so that the vertex correction vanishes after summing over $\bm{k}'$. Next, we plug Eq.~\eqref{eq:cooperon_abc} into Eq.~\eqref{eq:deltasigma} and find that $\delta\sigma=\delta\sigma_0 + \delta\sigma_s$, where $\delta\sigma_{0,s}$ are contributed by Cooperons $\phi_{\bm{k} ij}^{0,s}$, respectively. Since $\delta\sigma_0$ obeys the universal expression Eq.~\eqref{eq:deltasigma_2}, we only calculate $\delta\sigma_s$, whose expression is analogous to Eq.~\eqref{eq:deltasigma_n}.
\begin{equation}
    \delta\sigma_s^{\alpha\beta} = -\frac{e^2}{2 \pi\hbar}\int d\epsilon \frac{\partial n_{F}}{\partial \epsilon}[N_{1,s}^{\alpha\beta}+N_{2,s}^{\alpha\beta}+(N_{2,s}^{\alpha\beta})^{*}]\int \frac{d^{2} q}{(2\pi)^{2}} \frac{1}{D_s q^2 + \lambda_s\tau_0^{-1}}.
\end{equation}
Here,
\begin{align}
    N_{1,s}^{\alpha\beta} & = \frac{1}{2\pi\gamma\tau_0^2\Omega}\sum_{\bm k}\left(\hat{G}^{A}(-\bm k)\hat{v}^{\alpha}(-\bm k)\hat{G}^{R}(-\bm k)\right)_{ji'}\tau_{i'j'}^{y}\left(\hat{G}^{R}(\bm k)\hat{v}^{\beta}(\bm k)\hat{G}^{A}(\bm k)\right)_{ij'}\tau_{ij}^{y}\notag\\
    & = -\frac{1}{2\pi\gamma\tau_0^2\Omega}\sum_{\bm k} \tr\left[\hat{G}^{A}(\bm k)\hat{v}^{\alpha}(\bm k)\hat{G}^{R}(\bm k)\tau^{z}\hat{G}^{A}(\bm k)\hat{v}^{\beta}(\bm k)\hat{G}^{R}(\bm k)\tau^{z} \right]\notag\\
    & \approx -\frac{1}{\gamma\tau_0^2\Omega}\sum_{\bm k}\sum_{l=\pm}\delta(\epsilon- \epsilon_{\bm{k}l})v_{\bm{k}l}^{\alpha}v_{\bm{k}l}^{\beta} (n_{\bm{k}l}^{z})^2\tau_{\bm{k} l}^3\notag\\
    & = -D_{s}\delta^{\alpha\beta} .
\end{align}
To arrive at the second line, we used the time-reversal symmetry $\tau^x\hat{G}^{R/A}(\bm k)^{t}\tau^{x}= \hat{G}^{R/A}(-\bm k)$ and $\tau^x\hat{\bm v}(\bm k)^{t}\tau^{x}= -\hat{\bm v}(-\bm k)$. In the third line, we applied again the approximation $|\Delta_{\bm k}|\gg\tau_{0}^{-1}$ so that all propagators are in the same band $l$. 
\begin{align}
    N_{2,s}^{\alpha\beta} &= \frac{u_0^2}{2\pi\gamma\tau_0^2\Omega^2}\sum_{\bm k,\bm{k}'}\left(\hat{G}^{A}(-\bm k')\hat{v}^{\alpha}(-\bm k')\hat{G}^{R}(-\bm k')\hat{G}^{R}(-\bm k)\right)_{j'i'}\tau_{i'j}^{y}\left(\hat{G}^{R}(\bm k')\hat{G}^{R}(\bm k)\hat{v}^{\beta}(\bm k)\hat{G}^{A}(\bm k)\right)_{ij}\tau_{j'i}^{y}\notag\\
    & = -\frac{1}{\tau_0^3\Omega^2}\sum_{\bm k} \tr\left[\hat{G}^{A}(\bm{k}')\hat{v}^{\alpha}(\bm{k}')\hat{G}^{R}(\bm{k}')\hat{G}^{R}(\bm k)\tau^{z}\hat{G}^{A}(\bm k)\hat{v}^{\beta}(\bm k)\hat{G}^{R}(\bm k)\hat{G}^{R}(\bm{k}')\tau^{z} \right] \notag\\
    & \approx  -\frac{(2\pi)^2}{\tau_0^3\Omega^2}\sum_{\bm{k}l} \sum_{\bm{k}'l'}\delta(\epsilon- \epsilon_{\bm{k}'l'})\delta(\epsilon- \epsilon_{\bm{k}l})v_{\bm{k}' l'}^{\alpha}\bm n_{\bm{k}'l}^{z}\tau_{\bm{k}'l'}^2v_{\bm{k} l}^{\beta}\bm n_{\bm{k}l}^{z}\tau_{\bm{k}l}^2|\langle\bm{k}l |\bm{k}'l'\rangle|^2 \notag\\
    & = -\frac{(2\pi)^2}{2\tau_0^3\Omega^2}\sum_{\mu=0,x,y,z}\left(\sum_{\bm{k}l}\delta(\epsilon- \epsilon_{\bm{k}l})v_{\bm{k} l}^{\alpha} n_{\bm{k}l}^{z}n_{\bm{k}l}^{\mu}\tau_{\bm{k}l}^2\right)\left(\sum_{\bm{k}l}\delta(\epsilon- \epsilon_{\bm{k}l})v_{\bm{k} l}^{\beta} n_{\bm{k}l}^{z}n_{\bm{k}l}^{\mu}\tau_{\bm{k}l}^2\right)\notag\\
    & = 0.
\end{align}
In the second last equation, both terms in the bracket vanish because of $C_{3z}$ symmetry, $n_{C_{3z}\bm{k},l}^{\mu}=n_{\bm{k},l}^{\mu}$, $\tau_{C_{3z}\bm{k},l}=\tau_{\bm{k},l}$, and $\bm v_{C_{3z}\bm{k},l}=C_{3z}\bm v_{\bm{k},l}$. To summarize, the total conductance correction reads that
\begin{equation}
    \delta\sigma^{\alpha\beta}= -\frac{e^2}{2 \pi\gamma}\delta^{\alpha\beta}\left[D\int \frac{d^{2} q}{(2\pi)^{2}} \frac{1}{D q^2+\tau_{\phi}^{-1}}-D_{s}\int \frac{d^{2} q}{(2\pi)^{2}} \frac{1}{D_s q^2 + \lambda_s\tau_0^{-1}+\tau_{\phi}^{-1}}\right],
\end{equation}
where we include a decoherence rate $\tau_\phi^{-1}$ for both Cooperons. From this equation, it is straightforward to reproduce the weak-field magnetoconductance formula Eq.~(12) in the main text, which agrees to Eq.~\eqref{eq:deltasigma_B} with $\Xi=1$.

\subsection{Kramers intervalley coherent order in bilayer graphene}
In this section, we provide a concrete example for the weak antilocalization effect in Kramers-intervalley coherent (K-IVC) phases that preserve the Kramers time-reversal symmetry $\mathcal{T}_{K}=\tau^y\mathcal{K}$. We study the following K-IVC mean-field Hamiltonian of Bernal stacked bilayer graphene, $\hat{H}^{0}=\hat{H}_{\text{iso}}+\hat{H}_w$, where
\begin{align}\label{eq:h_bi}
    \hat{H}_{\text{iso}} &= h^{x}({\bm p})\sigma^{x} + h^{y}({\bm p})\tau^z\sigma^{y}+\Delta\tau^{y}\sigma^{y},\quad \bm h(\bm p)=(\frac{p_x^2-p_y^2}{2m}, \frac{p_xp_y}{m}),\\
    \hat{H}_{w} &= h_{w}^{x}({\bm p})\tau^z\sigma^{x} + h_w^{y}(\bm p)\sigma^{y},\quad \bm h_{w}(\bm p)=(v_3p_y, v_3p_x).
\end{align}
In this low-energy two-band (per valley) Hamiltonian, $\sigma^{i}$ act on the sublattice, or equivalently layer, degrees of freedom. $\hat{H}_{\text{iso}}$ has isotropic band dispersion, while $\hat{H}_w$ generates the trigonal warping. The intervalley-coherent order parameter $\tau^{y}\sigma^{y}$ breaks the microscopic time-reversal symmetry $\tau^x\mathcal{K}$ but preserves the Kramers time-reversal symmetry $\mathcal{T}_{K}$. Note that this order parameter anticommutes with $\hat{H}_{\text{iso}}$, $\{\tau^y\sigma^{y}, \hat{H}_{\textrm{iso}}\}=0$, and opens a band gap at the Dirac point. To avoid confusion, we emphasize that this is a fictitious model because K-IVC states have so far only been conjectured in magic-angle twisted multilayer graphene instead of untwisted graphene.

An important feature of this Bernal bilayer graphene model is that although $\hat{H}_{\text{iso}}$ breaks the valley $U(1)$ symmetry, it preserves a generalized valley symmetry, $[\hat{H}_{\text{iso}}, \tau^{x}\sigma^{x}]=0$. The quasiparticles can therefore be labeled by a flavor index $\xi=\tau^{x}\sigma^{x}=\pm 1$ and their wave functions are denoted as $\ket{\uparrow}$ and $\ket{\downarrow}$, respectively. We could further define for each flavor $\xi$ a generalized sublattice basis $\{|\tilde{A}\rangle,|\tilde{B}\rangle\}$ satisfying that $-\tau^y\sigma^{y}|\tilde{A}/\tilde{B}\rangle=\pm 1$, 
\begin{align}
    &|{\uparrow\tilde{A}}\rangle = \frac{1}{\sqrt{2}}\left(|K A\rangle + |{-K}B\rangle\right),\ |{\downarrow\tilde{A}}\rangle = \frac{1}{\sqrt{2}}\left(|KB\rangle - |{-K}A\rangle\right),\notag\\  
    &|{\uparrow\tilde{B}}\rangle = \frac{1}{\sqrt{2}}\left(|{KB}\rangle + |{-K}A\rangle \right),\ |{\downarrow\tilde{B}}\rangle = \frac{1}{\sqrt{2}}\left(|KA\rangle - |{-KB}\rangle\right).
\end{align}
Let us introduce Pauli matrices $\xi^{x,y,z}$ in the flavor space and $\rho^{x,y,z}$ in the generalized sublattice space, 
\begin{equation}
\begin{array}{lll}
\xi^{x}=-\tau^{z}\sigma^{x}, & \xi^{y}=\tau^{y}, & \xi^{z}=\tau^{x}\sigma^{x}, \\
\rho^{x}=\sigma^{x}, & \rho^{y} = \tau^y\sigma^z, & \rho^{z} = -\tau^y\sigma^{y}. 
\end{array}
\end{equation}
$\hat{H}_{\text{iso}}$ can be mapped to the Hamiltonian of bilayer graphene in a layer-polarizing field, with $\xi$ playing the role of valley,
\begin{equation}
    \hat{H}_{\text{iso}}(\bm p) = h^{x}({\bm p})\rho^{x} + h^{y}({\bm p})\xi\rho^{y} - \Delta\rho^{z}.
\end{equation}

\textit{Cooperons in the absence of trigonal warping and intervalley disorder scattering--}
To simplify the calculation, we first study the leading term of the mean-field Hamiltonian $\hat{H}_{\text{iso}}$ and assume the dominant source of disorder scattering to be the pseudospin-independent random potential from remote charge impurities, $\hat{V}(\bm r)=u(\bm r)\hat{1}$, with $\hat{1}$ a $4\times 4$ identity matrix and $\langle u(\bm r)u(\bm r')\rangle=u_0^2\delta(\bm r-\bm r')$. The trigonal warping and (spin-conserved) intervalley scattering disorders will be added perturbatively later. 

The self-consistent retarded and advanced Green's functions $\hat{G}^{R/A} = (\epsilon\hat{1}-\hat{H}_{\text{iso}}-\hat{\Sigma}^{R/A})^{-1}$ read that
\begin{align}
    &\hat{G}^{R/A}(\bm{k},\epsilon) = \frac{\epsilon \hat{1}+\hat{H}_{\text{iso}}}{(\epsilon\pm \frac{i\hbar}{2\tau})^2-\epsilon_{\bm k}^2} \label{eq:gf}, \\
    &\hat{\Sigma}^{R}(\bm{k},\epsilon) = -i\pi \gamma u_{0}^2\hat{1}+i\pi \gamma u_{0}^2\frac{\Delta}{\epsilon}\rho^{z}, \label{eq:selfenergy_kivc}
\end{align}
where we defined the band energy $\epsilon_{\bm k}=\sqrt{h(\bm k)^2+\Delta^2}$ and the quasiparticle relaxation time $\tau=\tau_s$,
\begin{equation}
    \tau_{s}^{-1}=\left(1+\frac{\Delta^2}{\epsilon^2}\right)\pi u_0^2\gamma,
\end{equation}
In this section, we focus on the Cooperons at $q=\omega=0$. So the Bethe-Salpeter equation becomes
\begin{equation}\label{eq:BS_bg}
    C_{\bm k m_1n_1,\bm k m_2n_2}=U_{m_{1}n_{1},m_{2}n_{2}}-\sum_{mn}\sum_{\bm pm'n'}U_{m_{1}n_{1},mn}G_{mm'}^{R}(\bm{p})G_{nn'}^{A}(-\bm{p})C_{\bm p m'n',\bm p m_2n_2},
\end{equation}
where the disorder scattering matrix elements are $U_{m_{1}n_{1},m_{2}n_{2}}=u_0^2\delta_{m_1,m_2}\delta_{n_1,n_2}$ if we neglect intervalley scattering ($u_v=0$) for the moment. Using the \textit{ansatz}, $C_{\bm k m_1m_2,\bm k n_1n_2}\equiv C_{m_1m_2,n_1n_2}$, we reduce the Bethe-Salpeter equation to a system of equations $\hat{K}_{0}\hat{C}=\hat{U}$,
\begin{equation}\label{eq:kernel_vs}
    \hat{K}_0 = \rho^0\otimes\rho^0-\frac{\pi u_0^2\gamma\tau_{s}}{2}\left[\rho^0\otimes\rho^0-\frac{\Delta}{\epsilon}(\rho^{z}\otimes\rho^0 + \rho^0\otimes\rho^{z})+\frac{\Delta^2}{\epsilon^2}\rho^{z}\otimes\rho^{z}+\frac{\epsilon^2-\Delta^2}{2\epsilon^2}(\rho^{x}\otimes\rho^{x}+\xi^z\rho^{y}\otimes\xi^{z}\rho^{y})\right]
\end{equation}
The eigenvalues and eigenvectors of $\hat{K}_0$ are summarized in Table~\ref{tab:gap}. For arbitrary $\Delta$, there are at least two degenerate low-frequency Cooperons in which two particle propagators carry opposite flavor indices. For either $\Delta/\epsilon\ll 1$ or $\Delta/\epsilon\rightarrow 1$, there are two extra low-frequency Cooperons in which two propagators carry the same flavor indices $\xi$, but the corresponding eigenfunctions of $\check{K}_0$ changes from $(|\xi\tilde{A},\xi\tilde{B}\rangle+|\xi\tilde{B},\xi\tilde{A}\rangle)/\sqrt{2}$ for $\Delta/\epsilon\ll 1$ to $|\xi\tilde{B},\xi\tilde{B}\rangle$ for $\Delta/\epsilon\rightarrow 1$ (Here $\xi=\uparrow,\downarrow$). We will focus on these six modes in the following discussions. All other modes always have a gap comparable to $\tau^{-1}$ and cannot lead to observable conductance corrections.

\begin{table}[]
    \centering
    \renewcommand{\arraystretch}{1.3}
    \caption{Cooperon gap in various valley-sublattice channels. Here the trigonal warping is not included and the disorder is chosen to be random scalar potentials independent of pseudospins. a) intra-flavor b) inter-flavor, $\alpha_{\pm}=(\epsilon\pm\Delta)/\sqrt{2(\epsilon^2+\Delta^2)}$.}
    \begin{tabular}{|>{\centering\arraybackslash}m{3.5 cm}|>{\centering\arraybackslash} m{2.7 cm}| >{\centering\arraybackslash}m{2.7 cm}|>{\centering\arraybackslash} m{2.7 cm}|>{\centering\arraybackslash} m{2.7 cm}|}
    \hline
       $\lambda$&$\frac{1}{\sqrt{2}}(|\tilde{A}\tilde{B}\rangle+|\tilde{B}\tilde{A}\rangle)$ & $\frac{1}{\sqrt{2}}(|\tilde{A}\tilde{B}\rangle-|\tilde{B}\tilde{A}\rangle)$ &$|\tilde{A}\tilde{A}\rangle$ & $|\tilde{B}\tilde{B}\rangle$ \\ \hline
       $\frac{1}{\sqrt{2}}(|\uparrow\uparrow\rangle\pm|\downarrow\downarrow\rangle)$& $\frac{2\Delta^2}{\epsilon^2+\Delta^2}$ & $1$ &$\frac{(\epsilon+\Delta)^2}{2(\epsilon^2+\Delta^2)}$ &$\frac{(\epsilon-\Delta)^2}{2(\epsilon^2+\Delta^2)}$ \\
       \hline
    \multicolumn{5}{c}{}\\
    \hline
        $\lambda$&$\alpha_{-}|\tilde{A}\tilde{A}\rangle+\alpha_{+}|\tilde{B}\tilde{B}\rangle$ & $\alpha_{+}|\tilde{A}\tilde{A}\rangle-\alpha_{-}|\tilde{B}\tilde{B}\rangle$ &$|\tilde{A}\tilde{B}\rangle$ & $|\tilde{B}\tilde{A}\rangle$ \\ 
        \hline
        $\frac{1}{\sqrt{2}}(|\uparrow\downarrow\rangle\pm|\downarrow\uparrow\rangle)$& $0$ & $1$ &$1-\frac{\epsilon^2-\Delta^2}{2(\epsilon^2+\Delta^2)}$&$1-\frac{\epsilon^2-\Delta^2}{2(\epsilon^2+\Delta^2)}$\\
       \hline
    \end{tabular}
    \label{tab:gap}
\end{table}
\begin{table}[]
    \centering
    \renewcommand{\arraystretch}{1.8}
    \caption{A list of the smallest eigenvalue of the Bethe-Salpeter kernel $\hat{K}$ in each flavor channel. We refer the full wave functions of these small-gap Cooperons to Table~\ref{tab:gap}. For the last two channels, the omitted generalized sublattice degree of freedom changes from $(|\tilde{A}\tilde{B}\rangle+|\tilde{B}\tilde{A}\rangle)/\sqrt{2}$ for $\Delta\ll \epsilon$ to $|\tilde{A}\tilde{A}\rangle$ for $\Delta\approx \epsilon$. $\tau_w^{-1}\approx 2(v_3p_{F})^2\tau$ and $\tau_{v}^{-1}$ are generated by trigonal warping and intervalley scattering, respectively.}
    \begin{tabular}{|c|c|c|c|c|}
        \hline
         $\lambda/\tau$ & $\frac{1}{\sqrt{2}}(|\uparrow\downarrow\rangle-|\downarrow\uparrow\rangle)$ & $\frac{1}{\sqrt{2}}(|\uparrow\downarrow\rangle + |\downarrow\uparrow\rangle)$& $\frac{1}{\sqrt{2}}(|\uparrow\uparrow\rangle - |\downarrow\downarrow\rangle)$ & $\frac{1}{\sqrt{2}}(|\uparrow\uparrow\rangle + |\downarrow\downarrow\rangle)$\\ \hline
        $\Delta\ll\epsilon$&$2\tau_v^{-1}$ & $\frac{\epsilon^2+\Delta^2}{\epsilon^2}\tau_w^{-1}+ \frac{\epsilon^2-\Delta^2}{\epsilon^2}\tau_v^{-1}$ & $\frac{2\Delta^2}{\epsilon^2+\Delta^2}\left[\tau^{-1}+\frac{\epsilon^2-\Delta^2}{2\epsilon^2}\left(\tau_w^{-1}-\tau_{v}^{-1}\right)\right]$ & $\frac{2\Delta^2}{\epsilon^2+\Delta^2}\tau^{-1}+\frac{\epsilon^2-\Delta^2}{\epsilon^2+\Delta^2}\left(\tau_w^{-1}+\tau_{v}^{-1}\right)$ \\ \hline
        $\Delta\approx\epsilon$& $2\tau_v^{-1}$ & $\frac{\epsilon^2+\Delta^2}{\epsilon^2}\tau_w^{-1} + \frac{\epsilon^2-\Delta^2}{\epsilon^2}\tau_v^{-1}$ & $\left(\frac{\epsilon+\Delta}{2\epsilon}\right)^2\left(\tau_w^{-1}+\frac{\epsilon^2-\Delta^2}{\epsilon^2+\Delta^2}\tau_v^{-1}\right)$ & $\left(\frac{\epsilon+\Delta}{2\epsilon}\right)^2\left(\tau_w^{-1}+\frac{3\epsilon^2+\Delta^2}{\epsilon^2+\Delta^2}\tau_v^{-1}\right)$ \\ \hline
    \end{tabular}
    \label{tab:fullgap}
\end{table}

\textit{Effects of trigonal warping --}The trigonal warping splits the doubly degenerate conduction bands into two bands with dispersion $\epsilon_{\pm}({\bm p}) = \left[\epsilon_{0}^2(\bm p)+h_{w}^2(\bm p)\pm f({\bm p})\right]^{1/2}$, where $f({\bm p})$ is the positive eigenvalues of the matrix 
\begin{equation}
    \hat{F} = 2\Delta h_{w}^{x}\xi^{x}\rho^{z} + 2\Delta h_{w}^y\xi^{y} + 2\bm h({\bm p})\cdot \bm h_w(\bm p) \xi^{x}\rho^{x}.
\end{equation} 
For simplicity, we consider the regime $|\bm h_{w}(\bm p)|\ll \hbar/\tau\ll$ the Fermi energy $\sim\sqrt{\Delta^2+p_F^{4}/4m^2}-\Delta$. The Green's function is modified as follows:
\begin{equation}
    \hat{G}^{R}\approx\frac{\left(\epsilon^2-\epsilon_{0}^2(\bm p)-h_w^2(\bm p) + \hat{F}(\bm p) \right)(\epsilon\hat{1}+\hat{H}_{\text{iso}})}{\left[(\epsilon+\frac{i}{2\tau})^2-\epsilon_{+}^2(\bm p)\right]\left[(\epsilon+\frac{i}{2\tau})^2-\epsilon_{-}^2(\bm p)\right]},
\end{equation}
One can show that the correction of the quasiparticle relaxation rate due to the trigonal warping is on the order of $O(h_w^2/\epsilon)$ and is therefore negligible. Plugging the above equation into the Bethe-Salpeter equation, Eq.~\eqref{eq:BS_bg}, we obtain $\hat{K}=\hat{K}_0+\hat{K}_w$,
\begin{equation}
    \hat{K}_w \approx 2\pi u_0^2\tau^3\frac{2}{(2\epsilon)^4}\int\frac{d^2p}{(2\pi)^2}\delta(\epsilon-\epsilon_{0}(\bm p))\left(f(\bm p)^2\hat{1}\otimes\hat{1}+\hat{F}(\bm p)\otimes\hat{F}(\bm p)\right)\cdot(\epsilon \hat{1}+ \hat{H}_{\text{iso}})\otimes(\epsilon \hat{1}+ \hat{H}_{\text{iso}}).
\end{equation}
By projecting $\hat{K}$ into the subspace of the six small-gap Cooperon modes of $\hat{K}_0$, we obtained the eigenvalues of $\hat{K}$ listed in Table~\ref{tab:fullgap} (set $\tau_v^{-1}=0$ for the moment). For any $\Delta$, there is always a gapless mode in the flavor-singlet channel, $({|\uparrow\downarrow\rangle}-{|\downarrow\uparrow\rangle})\otimes(\alpha_{-}|\tilde{A}\tilde{A}\rangle+\alpha_{+}|\tilde{B}\tilde{B}\rangle)/\sqrt{2}$. This wave function agrees with the general formula, Eq.~\eqref{eq:phi0}, with the self-energy Eq.~\eqref{eq:selfenergy_kivc} and $U_{T}=\tau^{y}$. In addition, the third mode in Table~\ref{tab:fullgap}, which is the intervalley triplet Cooperon in the original valley-sublattice basis, is gapless in the limit of $\Delta=0$ but acquires a gap $\sim 2\Delta^2/\epsilon^2$ for small $\Delta$ due to $\mathcal{T}_{O}$ symmetry breaking. Overall, there is only a single gapless Cooperon at finite $\Delta$ associated with the $\mathcal{T}_{K}$ symmetry and it will induce WAL and negative magnetoconductance according to Eq.~\eqref{eq:magneto}.

The results for $\Delta/\epsilon\approx 1$ can be derived in a simpler way. In this limit, quasiparticles in the conduction band have an approximately conserved quantum number $\rho^{z}\equiv -\tau^y\sigma^{y}=-1$. We can write down an effective Hamiltonian in the conductance band via the perturbation theory,
\begin{equation}
    \hat{H}_{\rho^{z}=-1} = \epsilon_{0}(\bm p)\xi^{0} + v_3(p_x\xi^{y} - p_y\xi^{x}) + \hat{V},
\end{equation}
where $\epsilon_0(\bm p) = \Delta+h^2(\bm p)/2\Delta$ is isotropic. This Hamiltonian equivalently describes a spinful particle with kinetic energy $h(\bm p)^2/2\Delta$ and Rashba spin-orbit coupling scattered by non-magnetic disorders, which is known to exhibit weak anti-localization \cite{iordanskii1994weak}.

\textit{Effects of intervalley disorder scattering--}
Here we consider the intervalley disorder $V_v=u_{vx}(\bm r)\tau^x+ u_{vy}(\bm r)\tau^y=u_{vx}(\bm r)\xi^{z}\rho^{x}+ u_{vy}(\bm r)\xi^{y}$, which could originate from atomically sharp defects. $\langle u_{vi}(\bm r)u_{vj}(\bm r')\rangle = u_{v}^2\delta_{i,j}\delta(\bm r-\bm r')$ implies the conservation of graphene crystal momentum after the disorder average. The self-energy is modified as follows,
\begin{align}
    \hat{\Sigma}^{R} &= -i\pi \gamma(u_{0}^2+2u_v^2)\hat{1}+i\pi \gamma u_{0}^2\frac{\Delta}{\epsilon}\rho^z.
\end{align}
The quasiparticle relaxation rate in Eq.~\eqref{eq:gf} becomes $\tau^{-1}=\tau_{s}^{-1}+\tau_{v}^{-1}$, where $\tau_{v}^{-1}=2\pi u_v^2\gamma \ll \tau_s^{-1}$.

In the Bethe-Salpeter equation, Eq.~\eqref{eq:BS_bg}, the disorder scattering matrix elements now include the contribution from the intervalley disorder scattering,
\begin{equation}\label{eq:vmatrixelement}
U_{i_{1}j_{1},i_{2}j_{2}}=u_0^2\delta_{i_1,i_2}\delta_{j_1,j_2}+u_v^2\left[(\tau^{x})_{i_1,i_2}(\tau^{x})_{j_1,j_2}+(\tau^{y})_{i_1,i_2}(\tau^{y})_{j_1,j_2}\right].    
\end{equation}
The Bethe-Salpeter kernel becomes $\hat{K} = \hat{K}_{0} + \hat{K}_{v}$,
\begin{align}
    \hat{K}_{v} = \left[\frac{\tau_s}{\tau_v} - \frac{\epsilon^2+\Delta^2}{2\epsilon^2}\frac{\tau_s}{\tau_v}\left(\xi^{z}\rho^{x}\otimes\xi^{z}\rho^{x}+\xi^{y}\otimes\xi^{y}\right)\right] \left( \rho^0\otimes\rho^0 -\hat{K}_{0} \right), 
\end{align}
where the first term in the square bracket comes from Eq.~\eqref{eq:kernel_vs} after replacing $\tau_{s}$ by $\tau\approx \tau_s(1-\tau_s/\tau_v)$. The second term is derived from the intervalley disorder scattering in Eq.~\eqref{eq:vmatrixelement} by using the relation $u_v^2/u_0^2=\left(1+\Delta^2/\epsilon^2\right)\tau_s/2\tau_v$. Projecting $\hat{K}$ into the subspace of the small-gap Cooperon modes of $\hat{K}_0$ mentioned in the previous section, we obtain new eigenvalues listed in Table~\ref{tab:fullgap} with $\tau_w^{-1}=0$ in this case. We see that intervalley scattering gaps both Cooperons in the flavor-singlet channel.
\\


Finally, let us include both trigonal warping and the intervalley scattering, $\hat{K} = \hat{K}_0+\hat{K}_w+\hat{K}_v$. The low-gap Cooperons and their gaps are summarized in Table~\ref{tab:fullgap}. The gap values in the limit of $\Delta\rightarrow 0$ are consistent with the well-known results of the normal bilayer graphene \cite{fal2007weak}: the first and third Cooperons in Table~\ref{tab:fullgap} respectively correspond to the valley-singlet mode with gap $2\tau_v^{-1}$ and the gapless valley-triplet mode, while the other two modes become the intravalley Cooperons with gap $\tau_w^{-1}+\tau_v^{-1}$ ($\tau_z^{-1}$ in Ref.~\cite{fal2007weak} is not included in our model). In the high-quality graphene devices with rare intervalley disorder scattering, $\tau_{v}^{-1}\ll \tau_{w}^{-1}$ is fulfilled, and the lowest Cooperon mode in the presence of a strong K-IVC order is the flavor-singlet mode in Table~\ref{tab:fullgap}, yielding a positive magnetoresistance in a weak out-of-plane magnetic field.

\end{document}